\documentclass[11pt,a4paper]{article}

\usepackage{geometry}

\usepackage{hyperref}
\usepackage{graphicx}
\usepackage{subcaption}
\usepackage{authblk}
\usepackage{amsfonts}
\usepackage{amsmath}
\usepackage{algorithm}
\usepackage{algpseudocode}
\usepackage{multirow}
\usepackage{graphics}
\usepackage{arydshln}
\usepackage[font=footnotesize, skip=0pt]{caption}

\providecommand{\keywords}[1]
{
	\small	
	\textbf{Keywords:} #1
}

\title{PBiLoss: Popularity-Aware Regularization to Improve Fairness in Graph-Based Recommender Systems}

\author{
	Mohammad Naeimi\thanks{\href{mailto:mohammad.naeimi@aut.ac.ir}{\texttt{mohammad.naeimi@aut.ac.ir}}} \space
    and
    Mostafa Haghir Chehreghani\thanks{\href{mailto:mostafa.chehreghani@aut.ac.ir}{\texttt{mostafa.chehreghani@aut.ac.ir}}} \\
	Department of Computer Engineering \\
	Amirkabir University of Technology (Tehran Polytechnic) \\
	Tehran, Iran
}

\date{}

\begin{document}

\maketitle

\vspace{-10mm}

\begin{abstract}
Recommender systems based on graph neural networks (GNNs) have been proved to perform well on user-item interactions. However, they commonly suffer from popularity bias---the tendency to over-recommend popular items---resulting in less personalization, unfair exposure and lower recommendation diversity. Current solutions address popularity bias through different stages of the recommendation pipeline, including pre-processing methods that may distort data distributions, in-processing approaches which can complicate optimization, and post-processing techniques that are limited in correcting bias already embedded in the learned representations.

To address these limitations, we propose PBiLoss, a novel regularization-based loss function designed to explicitly counteract popularity bias in graph-based recommenders. PBiLoss augments traditional training objectives by penalizing the model's inclination toward popular items, thereby encouraging the recommendation of less popular but potentially more personalized content. We introduce two sampling strategies---Popular Positive (PopPos) and Popular Negative (PopNeg)--- and explore two methods to distinguish popular items---one based on a fixed popularity threshold and another without any threshold---making the approach flexible and adaptive. Our proposed method is model-agnostic and can be seamlessly integrated into state-of-the-art graph-based frameworks such as LightGCN and its variants.

Extensive experiments carried out on datasets including Epinions, iFashion, and MovieLens highlight the advantages of the PBiLoss for enhancing fairness in recommendations, decreasing PRU and PRI by up to 10\%, compared to other baseline models, while maintaining accuracy and other standard metrics intact in the process.
\end{abstract}


	\section{Introduction}

Recommender systems play a critical role in modern digital platforms, improving user engagement and satisfaction across a wide range of applications, including social networks, streaming services, and e-commerce platforms. By leveraging large-scale user-item interaction data and auxiliary information, these systems aim to provide personalized recommendations that align with individual user preferences and behavioral patterns \cite{DBLP:reference/sp/2022rsh, DBLP:journals/csur/ZhangYST19}. Despite their effectiveness and widespread adoption, recommender systems suffer from a persistent fairness-related issue known as \emph{popularity bias}, where popular items---typically characterized by high interaction counts---are disproportionately recommended. As a result, long-tail items---although potentially highly relevant---receive limited exposure, reducing recommendation diversity, reinforcing existing popularity trends, and raising concerns regarding fairness, personalization, and provider exposure \cite{DBLP:conf/kdd/HajianBC16,DBLP:journals/umuai/AbdollahpouriAB20}. Popularity bias is particularly pronounced in graph-based recommender systems such as LightGCN \cite{DBLP:conf/sigir/0001DWLZ020}, which model user-item interactions as graphs and rely on message passing to propagate preference signals. While effective for capturing collaborative signals, graph propagation inherently amplifies dominant interaction patterns, creating a feedback loop that further boosts popular items while marginalizing less popular items \cite{DBLP:conf/icdm/RenW0GLF22}. This imbalance not only compromises the fairness of the recommendations but can also adversely affect user satisfaction by preventing the discovery of novel or niche content.

To mitigate this issue, prior work has explored several categories of solutions. Post-processing approaches, such as re-ranking methods \cite{DBLP:conf/flairs/AbdollahpouriBM19}, adjust recommendation lists after training to promote diversity, whereas re-weighting strategies \cite{DBLP:journals/jiis/GuptaKJ24} and gradient-based fairness interventions \cite{DBLP:conf/icdm/RenW0GLF22} modify the learning process itself. However, re-ranking methods operate independently of model training and may yield unstable fairness outcomes across different data distributions \cite{DBLP:journals/aim/SonboliBEM22}. In contrast, gradient-based approaches directly alter the optimization objective but often introduce complex loss landscapes that can impair convergence and generalization \cite{DBLP:journals/vldb/PitouraSK22, DBLP:journals/tors/GeLFTLXLXZ25}. These limitations motivate the need for fairness-aware techniques that are tightly integrated with model training while remaining stable and scalable. An alternative and increasingly adopted direction is to mitigate popularity bias through regularization of the learning objective. Regularization-based methods incorporate fairness constraints directly into the loss function, typically by penalizing disparities in exposure, representation, or prediction across user or item groups \cite{DBLP:journals/inffus/JinWZZDXP23, DBLP:journals/tist/ZhaoWLCAD25}. Prior work in this area includes fairness-aware matrix factorization \cite{DBLP:journals/corr/abs-2209-04394}, graph-based embedding regularization \cite{DBLP:conf/www/WuCSHWW21}, and loss-based corrections that either constrain item-group exposure \cite{kiswanto2018fairness} or equalize optimization progress across popular and unpopular items \cite{DBLP:journals/corr/abs-2410-04830}. While effective, many of these approaches depend on sensitive attributes, require careful hyperparameter tuning, or face scalability challenges in large-scale implicit feedback settings.

Building on this line of work, we propose \emph{Popularity Bias Loss} (\textbf{PBiLoss}), a regularization-based loss term tailored to mitigating popularity bias during training of graph-based recommenders. PBiLoss introduces an explicit penalty that discourages the over-representation of popular items in the learning objective and promotes more balanced exposure across the item spectrum. Because it operates as a modular loss component, PBiLoss can be seamlessly integrated into existing graph-based recommendation architectures without structural changes, embedding popularity-aware fairness directly into the learned representations while preserving recommendation accuracy. Our approach is further strengthened by the introduction of two sampling strategies tailored to the nuances of popularity bias. The \emph{Popular Positive} method treats popular items within the set of relevant (positive) interactions as biased samples, while the \emph{Popular Negative} method focuses on the over-representation of popular items within the irrelevant (negative) set. In addition, we propose two distinct methods for distinguishing popular items---one employing a fixed popularity threshold and one that operates without such an explicit threshold---thus offering flexibility and adaptability to varied data distributions and application contexts\footnote{Our implementation of PBiLoss is publicly available at \url{https://github.com/MhmdNmi/PBiLoss}.}.

Our main contributions in this work are summarized as follows:
\begin{itemize}
	\item \textbf{Novel loss function:} We propose PBiLoss, a regularization-based loss function that directly penalizes the over-representation of popular items during training. This integration of fairness into the learning objective mitigates popularity bias and promotes content diversity within recommender systems.
	
	\item \textbf{Innovative sampling strategies:} We introduce two sampling strategies---Popular Positive and Popular Negative---that target different aspects of popularity bias. These strategies allow for fine-grained control of exposure imbalance by treating biased samples distinctly during the training process.
	
	\item \textbf{Distinguishing popular items methods:} We propose two methods for distinguishing popular items: one that uses a fixed popularity threshold and another that does not impose a threshold. These two approaches enhance both the flexibility and the practical applicability of our method across diverse datasets.
	
	\item \textbf{Model-agnostic integration:} Our framework is designed to be flexible and easily integrable into existing pairwise recommendation architectures (e.g., LightGCN~\cite{DBLP:conf/sigir/0001DWLZ020}), without necessitating any structural modifications. This enables seamless adoption in various real-world scenarios.
	
	\item \textbf{Extensive empirical evaluation:} We demonstrate the effectiveness of PBiLoss through comprehensive experiments on real-world datasets including Epinions, iFashion, and MovieLens. The results reveal substantial improvements in fairness metrics, such as the Popularity-Rank correlation for Users (PRU) and Popularity-Rank correlation for Items (PRI), while preserving or enhancing recommendation accuracy.
	
\end{itemize}

The remainder of this paper is organized as follows. Section~\ref{sec:related_works} reviews related work on fairness in recommender systems and techniques for mitigating popularity bias. In Section~\ref{sec:preliminaries}, we outline the key preliminaries relevant to our study. Section~\ref{sec:methodology} describes our methodology, including the formulation of PBiLoss. Section~\ref{sec:experiments} presents our experimental results, and finally, Section~\ref{sec:conclusion} concludes with a discussion of our findings and directions for future work.


\section{Related work}
\label{sec:related_works}

The growing importance of fairness and the challenge of mitigating popularity bias in recommender systems have garnered significant attention due to their implications for diversity and user satisfaction~\cite{DBLP:conf/recsys/EkstrandBD19}. This section reviews existing research on techniques for mitigating popularity bias, achieving fairness in recommender systems, and graph-based recommendation methods.

\begin{table}
	\renewcommand{\arraystretch}{1.2}
	\centering
	\scriptsize
	\caption{Summary of representative fairness methods in recommender systems.}
	\label{tbl:literature}
	\begin{tabular}{p{2.5cm} p{6cm} p{6cm}}
		\hline
		\textbf{Method} & \textbf{Main idea \& pros} & \textbf{Limitations \& cons} \\
		\hline
		NISER \cite{DBLP:journals/corr/abs-1909-04276} &
		Normalizes item and session embeddings on a unit sphere. Simple to plug into SR-GNN-style models; improves long-tail items. &
		Designed for session setting; limited to scale normalization. Its applicability to generic CF graphs may require adaptation. \\
		\hline
		HetroFair \cite{DBLP:journals/ijon/GholinejadC26} &
		Combines fairness-aware attention with heterophily-aware feature weighting. Improves exposure of underrepresented items in heterophilous graphs.&
		More complex architecture; fairness terms need tuning. \\
		\hline
		PAAC \cite{DBLP:conf/kdd/Cai0WBSWZ024} &
		Aligns unpopular-item embeddings with popular ones and reweights contrastive loss. Gains in NDCG and long-tail coverage over LightGCN. &
		Contrastive training is costly and hyperparameter-sensitive. \\
		\hline
		PopDRL \cite{zhang2025relieving} &
		Uses debiasing representation objectives to downweight popularity signals without major architecture changes. &
		Distributionally robust training adds optimization overhead. \\
		\hline
		r-AdjNorm \cite{DBLP:conf/sigir/0002WLCZDWSLW22} &
		Adjusts neighborhood normalization to reduce impact of high-degree nodes. Plug-in normalization; small architectural changes. &
		Mainly handles topological bias; sensitive to hyperparameters. \\
		\hline
		APDA \cite{DBLP:conf/sigir/ZhouCDZZ023} &
		Uses inverse popularity scores and adaptive edge weights in message passing. Explicit, learnable popularity-aware weighting; keeps high-order signals. &
		Per-edge weights add parameters and computation. \\
		\hline
		PopGo \cite{DBLP:journals/tois/ZhangMZWC24} &
		Learns an explicit shortcut model and corrects interaction-level popularity shortcuts. Improves in-/out-of-distribution performance. &
		Requires an additional shortcut component and balancing. \\
		\hline
	\end{tabular}
\end{table}

\subsection{Graph neural networks as recommender systems}

Graph neural networks (GNNs) rapidly emerge as a cornerstone of graph-based machine learning, owing to their ability to capture complex relationships in structured data. The advent of Graph Convolutional Networks (GCN) \cite{DBLP:conf/iclr/KipfW17} marked a turning point by enabling efficient neighborhood aggregation through spectral convolutions. This breakthrough was followed by Graph Attention Networks (GAT) \cite{DBLP:conf/iclr/VelickovicCCRLB18}, which enhanced expressiveness by learning adaptive attention weights over neighboring nodes, and by GATv2 \cite{DBLP:conf/iclr/Brody0Y22}, which addresses the limitations of static attention mechanisms with a more flexible and dynamic formulation. More recently, TransGNN \cite{DBLP:journals/corr/abs-2401-01384} introduces an innovative approach that integrates strong transitivity relations to capture both local and global structural similarities, offering a deeper insight into graph topology. Collectively, these advancements exemplify the ongoing innovation and momentum in GNN research.

Early work on applying GNNs in recommender systems focused on adapting existing architectures to recommendation tasks. For example, Berg et al. \cite{DBLP:journals/corr/BergKW17} introduce Graph Convolutional Matrix Completion (GCMC), which employs a graph autoencoder to complete the user-item interaction matrix. Similarly, PinSage \cite{DBLP:conf/kdd/YingHCEHL18}, demonstrates the effectiveness of GNNs for large-scale recommendation tasks by leveraging the GraphSAGE algorithm \cite{DBLP:conf/nips/HamiltonYL17}.

Wang et al. \cite{DBLP:conf/sigir/Wang0WFC19} introduce Neural Graph Collaborative Filtering (NGCF), a model that captures higher-order connectivity by combining representations from multiple layers of the user-item bipartite graph. As one of the pioneering approaches in neural graph collaborative filtering, NGCF demonstrates superior performance compared to traditional methods, such as matrix factorization \cite{DBLP:journals/computer/KorenBV09} and neural collaborative filtering \cite{DBLP:conf/www/HeLZNHC17}. Sun et al. \cite{DBLP:conf/sigir/SunZGGTHMC20} show that simple aggregation operations (e.g., sum and mean) are insufficient for effectively modeling relational information in a node's neighborhood, and propose the Neighbor Interaction Aware Graph Convolutional Network (NIA-GCN) to address this limitation. Furthermore, Wang et al. \cite{DBLP:conf/sigir/WangJZ0XC20} develop Disentangled Graph Collaborative Filtering (DGCF), which accounts for users' independent preferences.

Chen et al. \cite{DBLP:conf/aaai/ChenWHZW20} simplify the model by removing nonlinear activation functions, demonstrating that this simplification enhances recommendation performance. A major advancement was achieved by He et al. with the introduction of LightGCN \cite{DBLP:conf/sigir/0001DWLZ020}. This model further streamlines the GNN architecture by eliminating both feature transformations and nonlinear activations, thereby focusing solely on neighborhood aggregation. LightGCN employs a simplified graph convolution mechanism to propagate representations in user-item graphs. This approach not only improves performance but also enhances efficiency, making it particularly suitable for large-scale recommendation scenarios.

Building upon the success of LightGCN, Mao et al. propose UltraGCN \cite{DBLP:conf/cikm/MaoZXLWH21}. UltraGCN further simplifies the model by eliminating explicit graph convolution operations and instead employing a constraint-based optimization approach that achieves comparable performance with enhanced efficiency. Specifically, UltraGCN forgoes the traditional message passing process by approximating an infinite-layer graph convolution using a constraint loss.


\subsection{Fairness in recommender systems}

Fairness in recommender systems is a multifaceted research area, addressing concerns such as equal representation, group fairness, and individual fairness. Graph-based recommender systems---exemplified by LightGCN \cite{DBLP:conf/sigir/0001DWLZ020}---achieve state-of-the-art performance in modeling complex user-item interactions. These methods leverage graph structures to propagate preferences and learn high-quality embeddings. However, their reliance on graph topology makes them susceptible to biases, as the inherent structure tends to amplify the influence of high-degree (popular) nodes \cite{DBLP:journals/umuai/DeldjooJBDZ24}.

Earlier frameworks such as FairRec \cite{DBLP:conf/www/PatroBGGC20} provide foundational definitions and debiasing techniques for item exposure fairness in recommendations, predating more recent methods. Yang and Stoyanovich \cite{DBLP:conf/ssdbm/YangS17} provide a comprehensive framework for the fairness-aware recommendation, highlighting the trade-offs between fairness and accuracy. Additionally, Steck \cite{DBLP:conf/recsys/Steck18} proposes a calibration-based approach to ensure that recommendations align with users' true preferences without overemphasizing popular items. 

In recent years, there has been a surge in research efforts addressing recommendation biases to achieve fairness \cite{DBLP:journals/tois/0007D0F0023}. With the widespread adoption of GNNs, there is growing social concern that GNN-based models may produce discriminatory outcomes \cite{DBLP:conf/www/DongLJL22}. Several studies explore mitigating biases in GNN-based recommender systems.

For example, NISER \cite{DBLP:journals/corr/abs-1909-04276} applies normalization operations on learned representations to mitigate popularity bias. FairGNN \cite{DBLP:conf/wsdm/DaiW21} employs an adversarial learning framework that incorporates fairness-aware constraints to ensure balanced representations of less popular items during training, thereby reducing GNN bias by leveraging both graph structure and limited sensitive attributes. Similarly, Zhang et al. \cite{DBLP:conf/www/0003MWSW24} propose a general debiasing framework for graph-based collaborative filtering, which utilizes adversarial graph dropout to reduce bias while preserving the graph structural integrity.

Zhang et al. \cite{DBLP:journals/tois/ZhangMZWC24} introduce PopGo, a debiasing strategy that quantifies and mitigates interaction-level popularity shortcuts in collaborative filtering models, thereby enhancing both in-distribution and out-of-distribution recommendation performance. Similarly, Gholinejad and Chehreghani \cite{DBLP:journals/corr/abs-2502-15699} propose an edge classification framework within graph neural networks that disentangles popularity bias from genuine item quality, promoting fairness by better representing high-quality, less popular items without compromising accuracy.

Recent work has begun exploring fairness and bias mitigation in recommender systems through the lens of large language models (LLMs). For example, UP5 \cite{DBLP:conf/eacl/HuaGXJLZ24} investigates user-side fairness for LLM-based recommendation by introducing counterfactually fair prompt mechanisms to promote equitable treatment across sensitive features in foundation model-based recommenders.

Evaluation frameworks tailored to LLM-based recommenders have also been proposed: FairEval \cite{DBLP:journals/corr/abs-2504-07801} introduces personality-aware fairness metrics for assessing disparities in recommendations generated LLMs, highlighting fairness gaps across demographic and psychological dimensions.

Additionally, FACTER \cite{DBLP:conf/icml/FayyaziKP25} integrates fairness-aware conformal thresholding with adaptive prompt engineering to reduce semantic and demographic bias in LLM-driven recommendation outputs. These studies showcase emerging directions using LLMs both as recommenders and as bias mitigators/evaluators, complementing classical fairness methods and motivating hybrid approaches.

Beyond recommender systems, recent studies in knowledge graph representation learning have explored bias-aware and structure-sensitive reasoning mechanisms. For example, abductive reasoning has been employed for multilingual entity alignment across multiple knowledge graphs, enabling robust alignment under heterogeneous and incomplete graph structures \cite{DBLP:journals/eaai/AkhtarLXCLH25}.

While notable progress has been made in promoting fairness in recommender systems, existing methods often fall short in addressing the complex and nuanced nature of popularity bias. Given the pervasive nature of these biases and the growing societal emphasis on equitable recommendation outcomes, significant opportunities for improvement remain. Consequently, developing methods that more effectively balance fairness with recommendation quality continues to be a critical and evolving research direction \cite{DBLP:journals/csur/WuSZXC23}.


\subsection{Popularity bias in recommender systems}

Popularity bias arises when recommender systems disproportionately favor popular items, limiting the visibility of niche content and reinforcing established popularity trends. This phenomenon has been extensively studied, with early approaches focusing on post-processing techniques such as re-ranking. For example, Abdollahpouri et al. \cite{DBLP:conf/flairs/AbdollahpouriBM19} introduce a personalized re-ranking framework that adjusts item rankings to balance relevance and exposure. However, since such methods are applied after recommendation generation, they are limited in their ability to address bias at its source.

More recent works explore debiasing strategies implemented during training. Chen et al. \cite{DBLP:conf/icde/ChenYNP0020} introduce a loss function designed to penalize over-reliance on popular items, thereby demonstrating the potential of model-level interventions. Zhao et al. \cite{DBLP:conf/sigir/0002WLCZDWSLW22} propose r-AdjNorm, a plug-in that adjusts neighborhood aggregation to mitigate popularity bias while preserving recommendation accuracy. Zhou et al. \cite{DBLP:conf/sigir/ZhouCDZZ023} introduce APDA, a graph-based method that combats popularity bias by computing inverse popularity scores and adaptively adjusting aggregation weights during GNN message passing.

HetroFair \cite{DBLP:journals/ijon/GholinejadC26} presents a fair GNN-based recommendation model that integrates fairness-aware attention and heterophily-aware feature weighting during message passing, thereby mitigating popularity bias and enhancing item-side fairness. Cai et al. \cite{DBLP:conf/kdd/Cai0WBSWZ024} propose Popularity-Aware Alignment and Contrast (PAAC), which mitigates popularity bias by using supervision from popular‐item embeddings to improve representations of unpopular items and re-weighting contrastive learning to reduce representation separation.

Building on these directions, several notable works further advance mitigation strategies and highlight complementary trade-offs. A broad survey \cite{DBLP:journals/umuai/KlimashevskaiaJET24} synthesizes recent detection and mitigation techniques and emphasizes the diversity of approaches (re-ranking, propensity correction, representation debiasing, and GNN aggregation fixes). Repurposing and industrial-scale strategies are proposed to increase exposure of long-tail content in production settings directly, showing practical online effects on exposure metrics and indicating that interventions can be engineered at multiple system levels \cite{DBLP:journals/corr/abs-2410-02776}.

At the model level, PopDRL \cite{zhang2025relieving}, a debiasing representation enhancement method, reports improvements by altering item/user embeddings to reduce popularity signals without heavy architectural changes. Parallel lines of work explore distributionally robust \cite{DBLP:conf/www/WangCL0SGFCW24} and linear-time \cite{DBLP:conf/www/ZhangXFXLPL24} GNN architectures to improve robustness and scalability when applying sophisticated debiasing in large graphs, showing that scalability concerns are being actively addressed. Finally, recent work started investigating the role of large language models (LLMs) as recommenders and as tools for measuring or mitigating popularity bias, opening a new axis of research \cite{DBLP:journals/corr/abs-2406-01285}.

While these approaches have demonstrated promise, they often rely on computationally intensive procedures (e.g., contrastive re-weighting, inverse-propensity estimation, or additional embedding supervision) and require intricate hyperparameter tuning or precomputations that can impede deployment at the industrial scale. Recent efforts towards linear-time GNNs and representation-level debiasing reduce some of those costs, but important trade-offs remain between effectiveness, interpretability, and scalability; thus, designing methods that are both computationally efficient and robust across datasets remains an open challenge \cite{DBLP:conf/www/ZhangXFXLPL24, zhang2025relieving}.



\section{Preliminaries}
\label{sec:preliminaries}

This section outlines the fundamental concepts that form the basis for understanding the context and objectives of our work. In particular, we review graph neural networks, recommender systems, and fairness, highlighting their key aspects and interconnections. We summarize frequently used notations in Table \ref{tbl:notations}.

\begin{table}
	\centering
	\caption{Notations and their descriptions.\label{tbl:notations}}
	\begin{tabular}{ l l }
		\hline
		Notation & Description \\
		\hline
		\(\mathcal{L}_{BPR}\) & The BPR loss \\
		\(T_{BPR}\) & Set of training triples sampled for BPR loss \\
		\(U\) & Set of training users \\
		\(I_{u}^{+}\) & Set of items relevant to user \(u\) (i.e., positive items) \\
		\(I \setminus I_{u}^{+}\) & Set of items not relevant to user \(u\) (i.e., negative items) \\
		\(\mathcal{L}_{PBi}\) & The proposed popularity bias loss \\
		\(T_{PBi}\) & Set of training triples sampled for PBiLoss\\
		\(I_{unpop}\) & Set of unpopular items \\
		\(I_{pop}\) & Set of popular items \\
		\(\alpha\) & The popularity threshold used in the fixed popularity threshold method \\
		\(w\) & A hyperparameter that controls the contribution of the PBiLoss \\
		\(I_{u,unpop}^{+}\) & Set of unpopular items relevant to user \(u\) \\
		\(I_{u,pop}^{+}\) & Set of popular items relevant to user \(u\) \\
		\(I_{\text{pop}} \setminus I_{u,\text{pop}}^{+}\) & Set of popular items not relevant to user \(u\) \\
		\hline
	\end{tabular}
\end{table}


\subsection{Graph neural networks}

Graph neural networks (GNNs) are deep learning models specifically designed to process and analyze graph-structured data. Their remarkable ability to capture complex relationships and dependencies in non-Euclidean data structures has garnered significant attention in recent years~\cite{DBLP:journals/tnn/WuPCLZY21}. Furthermore, GNNs have extended their impact beyond traditional graph datasets to domains such as recommender systems, where user-item interactions can be effectively represented and leveraged through graph-based architectures~\cite{DBLP:journals/natmi/Chehreghani22}.

GNNs operate on data comprising nodes (vertices) and edges (connections between nodes). The fundamental idea behind these models is to iteratively learn representations for nodes, edges, or entire graphs by aggregating and transforming information from neighboring nodes~\cite{DBLP:journals/tnn/WuPCLZY21}.

The primary objective of a graph neural network is to map each node to a vector in a continuous space, commonly referred to as the embedding vector or node representation. GNNs leverage a message passing mechanism, where each node transmits its current representation to its immediate neighbors. This process is typically composed of three main components:
\begin{itemize}
	\item \textbf{Message passing:} Nodes share their current state with their neighbors.
	\item \textbf{Aggregation:} The messages received from surrounding nodes are combined.
	\item \textbf{Update:} The node representation is updated based on the aggregated information.
\end{itemize}

The Graph Convolutional Network (GCN) is among the earliest and most popular graph neural networks, utilizing the message propagation mechanism defined in Equation~\ref{eq:gcn}.
\begin{equation}
	H^{(l+1)} = \sigma\!\left(\tilde{D}^{-\frac{1}{2}}\,\tilde{A}\,\tilde{D}^{-\frac{1}{2}}\,H^{(l)}\,W^{(l)}\right).\label{eq:gcn}
\end{equation}
In Equation~\ref{eq:gcn}, $D$ denotes the diagonal matrix of node degrees, where each diagonal element is defined as $D_{ii} = \sum_{j} A_{ij}$. The augmented matrices $\tilde{A} = A+I$ and $\tilde{D} = D+I$ incorporate self-loops, with $I \in \mathbb{R}^{(|U|+|I|) \times (|U|+|I|)}$ representing the identity matrix. Here, $\sigma(\cdot)$ is the activation function, $H^{(l)}$ is the embedding matrix (node representations) at layer $l$, and $W^{(l)}$ is the learnable weight matrix of the same layer. Graph neural networks have demonstrated significant promise in recommender systems by capturing complex user-item interactions and leveraging the inherent graph structure of recommendation data~\cite{DBLP:journals/tnn/WuPCLZY21}.

\subsection{Graph-based recommender systems}
In graph-based recommender systems, a set of users $(U = \{{u}_{1},{u}_{2},\ldots,{u}_{|U|}\})$ and a set of items $(I=\{{i}_{1},{i}_{2},\ldots,{i}_{|I|}\})$ are the set of nodes $(V = U \cup I)$ in a bipartite graph $(G=(V, E))$, in which edges $(E)$ are the implicit feedbacks (such as purchase, view, or click) between users and items. The user-item interaction matrix $(R \in \mathbb{R}^{|U|\times|I|})$ is defined such that if ${e}_{u,i} \in E$ then ${R}_{u,i}=1$ otherwise ${R}_{u,i}=0$. In graph $G$, there are no edges that connect two users or two items, so the graph adjacency matrix $(A)$, which is used during message propagation in the graph neural network, will be as follows:
\[
A = \setlength{\arraycolsep}{3pt} 
\renewcommand{\arraystretch}{1.5} 
\begin{bmatrix}
	{0}_{|U|\times|U|} & \hspace{1pt}\vrule\hspace{1pt} & R \\
	\hline
	R^{T} & \hspace{1pt}\vrule\hspace{1pt} & {0}_{|I|\times|I|} \\
\end{bmatrix}.
\]


\subsection{Fairness}

Fairness in recommender systems entails the equitable treatment of all stakeholders, including users and content providers. It aims to mitigate biases that can lead to discriminatory or unequal outcomes. In practice, recommender systems often amplify existing data biases by disproportionately favoring popular items (i.e., popularity bias) or underrepresenting minority groups and niche interests~\cite{DBLP:journals/umuai/DeldjooJBDZ24, DBLP:conf/sigir/Wu00DL22}. As these systems increasingly influence our daily decisions and experiences, ensuring fairness is critical to maintaining trust, promoting diversity, and avoiding unintended social consequences~\cite{DBLP:journals/tois/WangM00M23}.

Ensuring fairness is particularly challenging due to its multifaceted nature. For example, user fairness emphasizes delivering diverse, relevant, and personalized recommendations, while provider fairness seeks to guarantee equal exposure for items, especially those that might otherwise be overlooked. Achieving these objectives often requires balancing competing priorities, such as accuracy, diversity, and fairness~\cite{DBLP:conf/sigir/Wu00DL22}.

To address fairness concerns, researchers have introduced various techniques, including re-weighting training data, modifying model architectures, adjusting loss functions, and post-processing recommendation outputs. In this work, we propose the Popularity Bias Loss (PBiLoss), a novel loss function that mitigates popularity bias by reducing the over-reliance on popular items and promoting the exposure of less popular ones. This strategy not only enhances overall recommendation quality but also ensures a more equitable distribution of exposure across items.

Popularity bias is a pervasive challenge in recommender systems that warrants careful consideration. It occurs when a small subset of popular items disproportionately dominates recommendation lists, leading to reduced exposure for the majority of less popular items~\cite{DBLP:journals/corr/abs-2109-05677,DBLP:journals/corr/abs-2310-02961}.
Several factors contribute to popularity bias:
\begin{itemize}
	\item \textbf{Data imbalance:} Popular items tend to accumulate more interactions in the training data, resulting in richer representations in the model.
	\item \textbf{Feedback loops:} As popular items are recommended more frequently, they receive additional interactions, which further increases their prominence.
	\item \textbf{Cold start problem:} New or unique items suffer from a lack of historical data, making it difficult for them to achieve visibility.
\end{itemize}

Researchers have proposed a variety of metrics and scenarios to quantify popularity bias, including:
\begin{itemize}
	\item \textbf{Popularity lift:} The ratio of the frequency with which an item is recommended to its inherent popularity.
	\item \textbf{Long-tail coverage:} The proportion of less popular (long-tail) items that appear in recommendation lists.
	\item \textbf{Popularity-rank correlation:} The degree to which an item's popularity correlates with its predicted rank~\cite{DBLP:conf/wsdm/Zhu0ZZWC21}.
\end{itemize}


\subsection{Popularity bias in graph-based recommender systems}
\label{sec:pre_experiment}

Before introducing our proposed debiasing method, we empirically demonstrate that existing graph-based recommender systems (as exemplified by LightGCN) exhibit substantial popularity bias, compromising fairness in recommendation outcomes. This preliminary analysis highlights the need for a dedicated bias-mitigation strategy, such as PBiLoss.

\paragraph{Experimental setup}
We train a standard LightGCN model on the MovieLens dataset, which is used in this paper with the same configuration described in Section~\ref{sec:experiments}. After training, we generate top-10 recommendations for each user and report the metrics for popular (top 20\% by degree) and unpopular items.

\paragraph{Results and analysis}
Table~\ref{tbl:pre_experiment} reveals three key observations.
First, accuracy metrics (F1@10, NDCG@10, and MAP@10) are substantially higher for popular items compared to unpopular ones, indicating a strong performance imbalance. 
Second, the positive values of PRU and PRI show that item popularity is positively correlated with ranking positions, meaning that more popular items consistently receive higher ranks. 
Third, the ratio of the average rank of popular items to that of unpopular items is significantly below 1, further confirming that popular items are systematically ranked higher.
These observations show a performance difference of graph-based recommender systems between popular and unpopular items; hence, popularity bias undermines their overall performance.

\begin{table}
	\renewcommand{\arraystretch}{1.2}
	\centering
	\caption{Preliminary analysis of popularity bias in LightGCN on the MovieLens dataset. Popular items (top 20\% by interaction degree) achieve substantially higher accuracy than unpopular items, while positive PRU/PRI values and the low average rank ratio indicate a systematic preference toward popular items.\label{tbl:pre_experiment}}
	\begin{tabular}{ l | c c c }
		\hline
		\multirow{2}{*}{Metric} & \multicolumn{3}{c}{MovieLens dataset} \\
		\cline{2-4}
		& Popular & Unpopular & Overall \\
		\hline
		F1@10 $\uparrow$ & 0.2333 & 0.0337 & 0.2004 \\
		NDCG@10 $\uparrow$ & 0.3490 & 0.0453 & 0.3075 \\
		MAP@10 $\uparrow$ & 0.2038 & 0.0206 & 0.1286  \\
		\hline
		PRU $\downarrow$ &  &  & 0.5678 \\
		PRI $\downarrow$ &  &  & 0.8159 \\
		\hline
		Avg. rank (pop)/(unpop) & & & 0.4687 \\
		\hline
	\end{tabular}
\end{table}

These results confirm that LightGCN disproportionately promotes popular items while suppressing long-tail content. 
This bias can be attributed to the nature of graph-based message passing, where high-degree (popular) items receive more frequent updates, and to the BPR loss, which reinforces frequently observed user--item interactions. 
As a result, even relevant unpopular items are systematically under-ranked, leading to unfair exposure and reduced recommendation diversity.
Thus, a method that explicitly penalizes the over-representation of popular items during training--such as the PBiLoss we introduce next--is necessary to restore balanced exposure without sacrificing accuracy.



\section{Methodology} \label{sec:methodology}

In this section, we present a novel approach to mitigating popularity bias in recommender systems, with a particular focus on graph neural network (GNN)-based models. First, in Subsection \ref{subsec:BPR_PBi}, we review the BPR Loss and its relation to popularity bias. Next, in Subsection \ref{subsec:sample_triples}, we describe the sampling strategy for training triples used in the BPR Loss. Then, Subsection \ref{subsec:PBiLoss} introduces the proposed Popularity Bias Loss (PBiLoss), along with its specific sampling mechanism for training triples. Finally, Subsections \ref{subsec:algorithm} and \ref{subsec:time_complexity} present the algorithm and analyze the time complexity of the proposed method, respectively.

\subsection{Bayesian Personalized Ranking (BPR) Loss and popularity bias}
\label{subsec:BPR_PBi}

Recommender systems, particularly those utilizing graph neural networks with pairwise approaches, often rely on BPR loss to ensure that more relevant items are ranked higher than less relevant ones. However, because the BPR loss function inherently ranks items based on their overall relevance to a broad user base, it tends to introduce a popularity bias---popular items that appeal to many users dominate the recommendations, overshadowing less popular but relevant items. This bias reduces the diversity of recommendations and limits users' exposure to specific items.

Rendle et al. \cite{DBLP:conf/uai/RendleFGS09} introduced a general optimization measure for personalized rankings, namely the maximum posterior estimator, derived from a Bayesian analysis of predicting personalized rankings based on implicit feedback. The BPR loss is widely used to optimize model parameters in recommender systems by ensuring that items a user has interacted with are ranked higher than unseen items. Specifically, the loss function is defined as:
\begin{equation}
	\mathcal{L}_{BPR} = \sum_{(u,i,j) \in T} -\ln \sigma\Bigl(\tilde{y}_{ui} - \tilde{y}_{uj}\Bigr) + \beta \|\Theta\|^{2}, \label{eq:bpr_loss}
\end{equation}
where \(T\) denotes the training dataset, consisting of triples \((u, i, j)\) with \(u\) representing a user, \(i\) a positive item sample, and \(j\) a negative item sample. The hyperparameter \(\beta\) controls regularization to prevent overfitting, and \(\Theta\) represents the model parameters to be learned during optimization. The terms \(\tilde{y}_{ui}\) and \(\tilde{y}_{uj}\) are the relevance scores for items \(i\) and \(j\) for user \(u\), computed as the dot product of the corresponding user and item embeddings.

\subsection{Sampling training triples}
\label{subsec:sample_triples}

The sampling process in the BPR loss computation is a critical component of the optimization method, as it generates training samples that directly reflect the objective of personalized ranking. The BPR loss is formulated using implicit feedback data. Let \(U\) denote the set of all users and \(I\) denote the set of all items. The observed implicit feedback is represented as:
\[
S \subseteq U \times I,
\]
where an element \((u, i) \in S\) indicates that user \(u\) has interacted with item \(i\).
The goal of the sampling process is to create training triples \((u, i, j)\) such that:
\begin{itemize}
	\item \textbf{\(u\):} the sampled user,
	\item \textbf{\(i\):} a positive item that user \(u\) has interacted with,
	\item \textbf{\(j\):} a negative item that user \(u\) has not interacted with.
\end{itemize}
The set of training triples, denoted by \(T_{BPR}\), is defined as:
\begin{equation}
	T_{BPR} := \{ (u,i,j) \mid u \in U ,\; i \in I_{u}^{+},\; j \in I \setminus I_{u}^{+} \}. \label{eq:bprtriples}
\end{equation}
In Equation \ref{eq:bprtriples}, \(U\) is the set of training users, \(I_{u}^{+}\) denotes the set of items that user \(u\) has interacted with (i.e., positive items), and \(I \setminus I_{u}^{+}\) denotes the set of items that user \(u\) has not interacted with (i.e., negative items).

The sampling process is executed once at the beginning of each training epoch, and the resulting set of training triples is used to compute the BPR loss and train the model during that epoch. The size of the sampled training set is treated as a hyperparameter; in our experiments, it is set equal to the number of user-item interactions in the training dataset. For each training instance in the sampled set, the following process is executed:
\begin{enumerate}
	\item A user \( u \) is randomly selected from the complete set of users \( (u \in U) \).
	\item A positive item \( i \) is randomly selected from the set of items with which user \( u \) has interacted \( (i \in I_{u}^{+}) \).
	\item A negative item \( j \) is randomly selected from the set of items with which user \( u \) has not interacted \( (j \in I \setminus I_{u}^{+}) \).
\end{enumerate}
For each training triple \((u, i, j)\), the BPR loss is computed as follows:
\begin{equation}
	\mathcal{L}_{BPR} = -\ln \sigma\Bigl(\tilde{y}_{ui} - \tilde{y}_{uj}\Bigr), \label{eq:bpr_sample}
\end{equation}
where in Equation \eqref{eq:bpr_sample} \(\tilde{y}_{ui}\) denotes the predicted preference of user \(u\) for item \(i\), and \(\tilde{y}_{uj}\) denotes the predicted preference for item \(j\). The function \(\sigma\) is the logistic sigmoid function, which is defined as:
\begin{equation}
	\sigma(x) = \frac{1}{1 + e^{-x}}. \label{eq:sigmoid}
\end{equation}

Model parameters are updated to minimize the loss, typically using stochastic gradient descent. The sampling process ensures that the model learns to rank positive items higher than negative items for each user. By repeatedly sampling and updating these triples, the model develops the ability to create personalized rankings in which items with which the user is likely to interact are ranked higher than those with which they are unlikely to engage.
This sampling approach is efficient because it does not require explicit negative feedback, which is often unavailable in real-world scenarios. Instead, unobserved interactions are treated as implicit negative feedback, enabling the model to learn from both observed and unobserved user-item interactions.


\subsection{Our proposed Popularity Bias Loss (PBiLoss)}
\label{subsec:PBiLoss}

To address the problem of popularity bias, we propose the Popularity Bias Loss (PBiLoss), which is designed to boost the ranking of unpopular but relevant items while demoting the ranking of popular items---particularly for users for whom these items are less relevant. Item popularity can be computed in various ways; a common approach is to use the number of interactions, i.e., the node degree of the item \cite{DBLP:journals/umuai/KlimashevskaiaJET24}.

\subsubsection{Distinguishing popular items}

In this research, each item's popularity was quantified by its node degree. Two methods were introduced to distinguish and differentiate between popular and unpopular items:
\begin{enumerate}
	\item
	\textbf{Fixed popularity threshold:} In this method, we introduce a fixed hyperparameter \(\alpha\) as the popularity threshold. An item \(i\) is deemed popular if its node degree, \(degree(i)\), is greater than or equal to \(\alpha\); otherwise, it is considered unpopular. Accordingly, when sampling:
	a popular item is selected uniformly at random from the set of items with \(degree(i) \geq \alpha\), and an unpopular item is selected uniformly at random from the set of items with \(degree(i) < \alpha\).
	
	Here, "uniformly at random" means that each item within the respective set has an equal chance of being selected. This method is formally represented as:
	\begin{equation}
		popular(i) =
		\begin{cases}
			\textnormal{True}, & \textnormal{if } degree(i) \geq \alpha, \\
			\textnormal{False}, & \textnormal{if } degree(i) < \alpha.
		\end{cases} \label{eq:fpopt}
	\end{equation}
	In Equation \eqref{eq:fpopt}, the function \(popular(i)\) determines whether item \(i\) is popular. The set of popular items consists of all items for which \(popular(i)\) is True, while the set of unpopular items comprises those for which \(popular(i)\) is False.
	
	\item
	\textbf{No popularity threshold:} In this alternative method, no fixed threshold is required. Instead, sampling probabilities are assigned individually to each item based on its popularity:
	When sampling a popular item from a set \(I\), the item's popularity is used directly as its selection probability. Conversely, when sampling an unpopular item, the inverse of the item's popularity is used.
	
	This approach ensures that the more popular an item is, the higher its chance of being selected as a popular sample. In contrast, less popular items are more likely to be chosen as unpopular samples. The corresponding sampling probabilities are defined as:
	\begin{equation}
		P_{pop}(i,I) = \frac{degree(i)}{\sum_{j \in I} degree(j)}, \label{eq:ppop}
	\end{equation}
	\begin{equation}
		P_{unpop}(i,I) = \frac{\frac{1}{degree(i)}}{\sum_{j \in I} \frac{1}{degree(j)}}. \label{eq:punpop}
	\end{equation}
	In Equation \eqref{eq:ppop}, \(P_{pop}(i,I)\) denotes the probability of sampling item \(i\) as a popular item from the set \(I\). Similarly, Equation \eqref{eq:punpop} defines \(P_{unpop}(i,I)\) as the probability of sampling item \(i\) as an unpopular item. Here, \(degree(i)\) represents the node degree of item \(i\), and \(I\) is the set of available items. Thus, when sampling items:
	\begin{itemize}
		\item A popular item is selected according to the distribution \(P_{pop}\).
		\item An unpopular item is selected according to the distribution \(P_{unpop}\).
	\end{itemize}
	We acknowledge that under this method, a sampled pair may occasionally consist of a less popular item as popular and a more popular item as unpopular, which by itself could reinforce popularity bias. However, our approach is defined by statistical intuition: the model is optimized over a huge number of sampled pairs, rather than a single instance. While some individual updates may appear unfavorable, the expected gradient across all samples shifts the model toward fair treatment of unpopular items. Moreover, we provide the fixed threshold variant as a stricter control, and both methods yield consistent improvements, demonstrating the robustness of our methodology.
\end{enumerate}
Note that here, the degree is merely one of many node centrality measures;
alternative centrality notions may also be employed~\cite{DBLP:journals/cj/ChehreghaniBA18,DBLP:conf/cikm/ChehreghaniBA19}.
The final formula for calculating PBiLoss is obtained by integrating the proposed popularity bias component into the standard BPR loss framework:
\begin{itemize}
	\item The BPR loss is a pairwise ranking loss that drives the model to rank relevant items higher than irrelevant ones.
	\item By leveraging the BPR framework, we can further encourage the model to rank unpopular items higher and demote the ranking of popular items.
\end{itemize}


\subsubsection{Sampling training triples in PBiLoss}

As mentioned, the design of PBiLoss is inspired by the structure of the BPR loss. Consequently, its formulation follows Equation~\ref{eq:bpr_loss}, with a modified sampling strategy that explicitly takes item popularity into account. In particular, the sampling process for PBiLoss generalizes the structure defined in Equation~\ref{eq:bprtriples} as follows:
\begin{equation}
	T_{PBi} := \{(u,i,j) \mid u \in U,\; i \in I_{unpop},\; j \in I_{pop} \}. \label{eq:pbtriples}
\end{equation}
In this context, \(I_{unpop}\) and \(I_{pop}\) denote the sets of unpopular and popular items, respectively. Two sampling strategies have been investigated within this framework: Popular Positive (PopPos) and Popular Negative (PopNeg). The PopPos strategy focuses on sampling pairs among relevant items---distinguishing between popular and unpopular items---to promote the ranking of relevant but under-represented items. Conversely, the PopNeg strategy addresses the ranking of negative (irrelevant) items by emphasizing the demotion of popular items that are not relevant to the user:

\begin{enumerate}
	\item\textbf{Popular Positive (PopPos).}
	This strategy emphasizes positive (i.e., relevant) items. The intuition is that popular items tend to receive higher rankings under the BPR loss because many users interact with them. By incorporating the PBiLoss on these items, the system is encouraged to elevate the ranking of less popular but still relevant items. Specifically, the PBiLoss is computed between pairs of positive items---one from the unpopular set and one from the popular set---thus allowing the model to emphasize those underrated items that deserve higher ranks and, in turn, enhancing recommendation diversity while maintaining relevance.
	
	In Popular Positive with a fixed popularity threshold, the dataset is first partitioned into two groups---popular items, denoted by $I_{\text{pop}}$, and unpopular items, denoted by $I_{\text{unpop}}$---as defined in Equation~\ref{eq:fpopt}. Sampling is then performed based on item popularity:
	\begin{equation}
		T_{PBi-PopPos} := \{ (u, i, j) \mid u \in U,\; i \in I_{u,unpop}^{+},\; j \in I_{u,pop}^{+} \}. \label{eq:pbi_poppos_fpt}
	\end{equation}
	Here, \(U\) represents the set of users, \(I_{u,unpop}^{+}\) denotes the set of unpopular items relevant to user \(u\), and \(I_{u,pop}^{+}\) denotes the set of popular items relevant to user \(u\).

	In Popular Positive with no popularity threshold, each user's set of relevant items is sampled using different probabilities defined by Equations~\ref{eq:ppop} and~\ref{eq:punpop}:
	\begin{equation}
		T_{PBi-PopPos} := \{(u,i,j) \mid u \in U,\; i,j \in I_{u}^{+},\; i \sim P_{unpop}(\cdot, I_{u}^{+}),\; j \sim P_{pop}(\cdot, I_{u}^{+}) \}. \label{eq:pbi_poppos_npt}
	\end{equation}
	
	In Equation~\ref{eq:pbi_poppos_npt}, \(U\) denotes the set of training users, and \(I_{u}^{+}\) represents the set of items relevant to user \(u\). The notation 
	\[
	i \sim P_{unpop}(\cdot, I_{u}^{+})
	\]
	indicates that item \(i\) is sampled from \(I_{u}^{+}\) using the probability distribution defined in Equation~\ref{eq:punpop}. Similarly, 
	\[
	j \sim P_{pop}(\cdot, I_{u}^{+})
	\]
	denotes that item \(j\) is sampled from the same set (\(I_{u}^{+}\)) using the probability distribution given in Equation~\ref{eq:ppop}.

	\item
	\textbf{Popular Negative (PopNeg).} This strategy extends the concept of PBiLoss to negative (i.e., irrelevant) items. In many recommender systems, popular items tend to appear higher in the ranked list, regardless of their relevance to a particular user. To counteract this tendency, the popular negative strategy places additional emphasis on lowering the rankings of popular items that are deemed irrelevant. In this approach, PBiLoss is computed between positive (relevant) unpopular items and negative (irrelevant) popular items, thereby boosting the ranking of relevant unpopular items while demoting popular items with which a user has not interacted.
	
	In the Popular Negative strategy with a fixed popularity threshold, the dataset is first partitioned into two groups: popular items, denoted by \(I_{\text{pop}}\), and unpopular items, denoted by \(I_{\text{unpop}}\), as defined in Equation~\ref{eq:fpopt}. Sampling is then performed based on item popularity:
	\begin{equation}
		T_{PBi-PopNeg} := \{(u,i,j) \mid u \in U,\; i \in I_{u,\text{unpop}}^{+},\; j \in I_{\text{pop}} \setminus I_{u,\text{pop}}^{+}\}. \label{eq:pbi_popneg_fpt}
	\end{equation}
	In Equation~\ref{eq:pbi_popneg_fpt}, \(U\) denotes the set of training users, \(I_{u,\text{unpop}}^{+}\) represents the set of unpopular items with which user \(u\) has interacted, and \(I_{\text{pop}} \setminus I_{u,\text{pop}}^{+}\) indicates the set of popular items with which user \(u\) has not interacted.

	In Popular Negative with no popularity threshold---similar to Popular Positive---sampling is performed over both the items that user \(u\) has interacted with and those that user \(u\) has not. Specifically, the sampling probabilities for each item are determined according to Equations~\ref{eq:punpop} and~\ref{eq:ppop} as follows:
	\begin{equation}
		T_{PBi-PopNeg} := \{ (u, i, j) \mid u \in U,\; i \in I_{u}^{+},\; j \in I \setminus I_{u}^{+},\; i \sim P_{unpop}(\cdot, I_{u}^{+}),\; j \sim P_{pop}(\cdot, I \setminus I_{u}^{+}) \}. \label{eq:pbi_popneg_npt}
	\end{equation}
	
	In Equation~\ref{eq:pbi_popneg_npt}, \(U\) denotes the set of training users, and \(I_{u}^{+}\) represents the set of items with which user \(u\) has interacted. The notation
	\[
	i \sim P_{unpop}(\cdot, I_{u}^{+})
	\]    
	indicates that item \(i\) is sampled from \(I_{u}^{+}\) using the probability distribution defined in Equation~\ref{eq:punpop}, while   
	\[
	j \sim P_{pop}(\cdot, I \setminus I_{u}^{+})
	\]
	denotes that item \(j\) is sampled from the complement set \(I \setminus I_{u}^{+}\) (i.e., items that user \(u\) has not interacted with) using the probability distribution given in Equation~\ref{eq:ppop}.
	
\end{enumerate}

Both the Popular Positive and Popular Negative strategies address popularity bias by redistributing item rankings based on both popularity and relevance. This adjustment effectively mitigates the over-representation of popular items in the recommendations. Either (or both) of these strategies can be used to compute the PBiLoss. We then combine the PBiLoss with the standard BPR loss during training as follows:
\begin{equation}
	\mathcal{L}_{total} = \mathcal{L}_{BPR} + w \cdot \mathcal{L}_{PBi}, \label{eq:total_loss}
\end{equation}
where \(\mathcal{L}_{BPR}\) represents the standard BPR loss, \(\mathcal{L}_{PBi}\) denotes the proposed popularity bias loss (PBiLoss), and \(w\) is a hyperparameter that controls the contribution of the PBiLoss in the overall training objective.


\subsection{Algorithm}
\label{subsec:algorithm}

\begin{algorithm}
	\footnotesize
	\caption{\footnotesize The proposed PBiLoss method training process for the LightGCN model.\label{alg:train_pbiloss}}
	\begin{algorithmic}[1]
		\State \textbf{Input:} \label{lin:inputs}
		\Statex \quad - Graph $G(V,E)$,
		\Statex \quad - number of LightGCN (or any other used GNN) layers ${n}_{L}$,
		\Statex \quad - number of epochs ${n}_{epochs}$,
		\Statex \quad - PBiLoss strategy ${strategy}_{PBi}$,
		\Statex \quad - PBiLoss weight $w$,
		\Statex \quad - Popularity Threshold $\alpha$.
		\State \textbf{Output:} Trained model embedding matrix $\Theta(X)$. \label{lin:outputs}
		\State $X \gets$ Xavier initialization \label{lin:init_model}
		\Comment\textsf{{Initializing input features of LightGCN (or any other used GNN)}}
		\For{$epoch=0$ to ${n}_{epochs}$}
		\State  $H \gets \Theta(X)$ \label{lin:calc_H}
		\Comment\textsf{Calculating embedding matrix}
		\State ${\mathcal{L}}_{total}, {\mathcal{L}}_{PBi}, {\mathcal{L}}_{BPR} \gets 0, 0, 0$ \label{lin:init_loss}
		\Comment\textsf{Initializing Loss}
		\State $T_{BPR} \gets$ BPRSampling($U,I,{I}^{+}$) \label{lin:BPRSampling}
		\Comment\textsf{$T_{BPR} := \{(u,i,j) \mid u \in U ,\; i \in I_{u}^{+} ,\; j \in I \setminus I_{u}^{+}\}$}
		\For{$(u,i,j) \in T_{BPR}$}
		\Comment\textsf{$T_{BPR}$ is the sampled triples for BPR Loss calculation}
		\State $\tilde{y}_{ui} \gets H[u] \cdot H[i]$ \label{lin:bpr_pos_score}
		\State $\tilde{y}_{uj} \gets H[u] \cdot H[j]$ \label{lin:bpr_neg_score}
		\State ${\mathcal{L}}_{BPR} \gets {\mathcal{L}}_{BPR} + (- \ln\sigma(\tilde{y}_{ui}-\tilde{y}_{uj})+ \beta ||X||^2)$ \label{lin:bpr_loss}
		\EndFor
		\If{${strategy}_{PBi}=="PopPos"$} \label{lin:start_pbisample}
		\If{$\alpha>0$}
		\State $T_{PBi} \gets$ PBiSampling\_PopPosT($U,I,{I}^{+},\alpha$)\label{lin:pbi_poppos_fpt}
		\Comment\textsf{Sampling as in Equation \ref{eq:pbi_poppos_fpt}}
		\Else
		\State $T_{PBi} \gets$ PBiSampling\_PopPosN($U,I,{I}^{+}$)\label{lin:pbi_poppos_npt}
		\Comment\textsf{Sampling as in Equation \ref{eq:pbi_poppos_npt}}
		\EndIf
		\ElsIf{${strategy}_{PBi}=="PopNeg"$}
		\If{$\alpha>0$}
		\State $T_{PBi} \gets$ PBiSampling\_PopNegT($U,I,{I}^{+},\alpha$)\label{lin:pbi_popneg_fpt}
		\Comment\textsf{Sampling as in Equation \ref{eq:pbi_popneg_fpt}}
		\Else
		\State $T_{PBi} \gets$ PBiSampling\_PopNegN($U,I,{I}^{+}$)\label{lin:pbi_popneg_npt}
		\Comment\textsf{Sampling as in Equation \ref{eq:pbi_popneg_npt}}
		\EndIf
		\EndIf \label{lin:end_pbisample}
		\For{$(u,i,j) \in T_{PBi}$}
		\Comment\textsf{$T_{PBi}$ is the sampled triples for PBiLoss calculation}
		\State $\tilde{y}_{ui} \gets H[u] \cdot H[i]$ \label{lin:pbi_unpop_score}
		\State $\tilde{y}_{uj} \gets H[u] \cdot H[j]$ \label{lin:pbi_pop_score}
		\State ${\mathcal{L}}_{PBi} \gets {\mathcal{L}}_{PBi} + (- \ln\sigma(\tilde{y}_{ui}-\tilde{y}_{uj}))$ \label{lin:pbi_loss}
		\EndFor \label{lin:end_pbi_for}
		\State ${\mathcal{L}}_{total} \gets {\mathcal{L}}_{BPR} + w \cdot {\mathcal{L}}_{PBi}$ \label{lin:total_loss}
		\State \textbf{Backpropagate and update} model parameters using gradient descent on ${\mathcal{L}}_{total}$ \label{lin:backpropagate}
		\EndFor
		\State \textbf{return} $\Theta(X)$
	\end{algorithmic}
\end{algorithm}

We present the training process of the proposed method (for the LightGCN model~\cite{DBLP:conf/sigir/0001DWLZ020}), detailing the algorithmic steps executed during each training epoch.
Algorithm~\ref{alg:train_pbiloss} outlines the complete training process of our approach.
As indicated in Line~\ref{lin:inputs}, the algorithm accepts as input an input graph \(G(V, E)\), the number of layers \(n_L\) for LightGCN (or any other employed GNN model), the total number of training epochs \(n_{epochs}\), the selected PBiLoss strategy \({strategy}_{PBi}\), the PBiLoss weight \(w\), and, when applicable, a popularity threshold \(\alpha\) for fixed threshold method. The algorithm produces as output the trained embedding matrix \(\Theta(X)\) for the LightGCN model, as shown in Line~\ref{lin:outputs}.

Line~\ref{lin:init_model} initializes the model parameters using the Xavier weight initialization method \cite{DBLP:journals/jmlr/GlorotB10}. Furthermore, as shown in Line~\ref{lin:calc_H}, the representation matrix is computed at the beginning of each training epoch. Line~\ref{lin:init_loss} indicates the initialization of the loss, and Line~\ref{lin:backpropagate} demonstrates the backpropagation step during which the loss is computed and the model parameters are updated. Note that these operations can also be performed in multiple batches.

Batching divides the training process into smaller subsets, thereby enhancing both efficiency and scalability. Within each epoch, the model processes multiple iterations, during which the loss for each batch is computed and used to update the model parameters. The BPRSampling function (Line~\ref{lin:BPRSampling}) handles the sampling operation necessary for calculating the BPR loss \cite{DBLP:conf/uai/RendleFGS09}. Specifically, in Lines~\ref{lin:bpr_pos_score} and \ref{lin:bpr_neg_score}, the predicted relevance scores for positive and negative items are computed for each user, while Line~\ref{lin:bpr_loss} calculates the BPR loss.

Also, the functions 
PBiSampling\_PopPosT, PBiSampling\_PopPosN, PBiSampling\_PopNegT, and PBiSampling\_PopNegN 
(Lines~\ref{lin:start_pbisample} to \ref{lin:end_pbisample}) perform the sampling operations for PBiLoss. 
Specifically, these functions implement the sampling processes for: 
(i) popular positive with a fixed popularity threshold, 
(ii) popular positive with no popularity threshold, 
(iii) popular negative with a fixed popularity threshold, and 
(iv) popular negative with no popularity threshold, 
respectively, and are invoked as required.
In Lines~\ref{lin:pbi_unpop_score} and \ref{lin:pbi_pop_score}, the model computes the user's predicted relevance scores for unpopular and popular items, respectively, and Line~\ref{lin:pbi_loss} then calculates the PBiLoss. Finally, Line~\ref{lin:total_loss} combines the BPR loss and PBiLoss using the weight \( w \) to calculate the total loss, and Line~\ref{lin:backpropagate} performs backpropagation by computing the derivative of the total loss.


\subsection{Time complexity analysis}
\label{subsec:time_complexity}

Recommender systems must achieve both efficiency and accuracy---a balance that is often challenging to strike. In this section, we evaluate the time complexity of our proposed algorithm. Notably, if the components in Lines~\ref{lin:start_pbisample} to \ref{lin:end_pbi_for} of Algorithm~\ref{alg:train_pbiloss} are omitted, our model essentially reduces to LightGCN \cite{DBLP:conf/sigir/0001DWLZ020} with the BPR Loss \cite{DBLP:conf/uai/RendleFGS09}.

The streamlined architecture of LightGCN distinguishes it from more complex models such as SGL-ED \cite{DBLP:conf/sigir/WuWF0CLX21}. This simplicity contributes significantly to its effectiveness and efficiency. By eliminating extraneous components---such as transformation matrices and nonlinear activators---LightGCN not only achieves lower computational complexity but also often delivers performance that rivals or exceeds that of more complex GNN models.

The graph convolution step in LightGCN scales linearly with the number of edges \(|E|\), the number of layers \(L\), and the embedding size \(d\). Consequently, the graph convolution, with a time complexity of \(O(2|E|Ld)\) (where \(L\) and \(d\) are fixed), is the primary computational cost---contributing significantly to LightGCN's practical efficiency.

In our proposed method, the PBiLoss is computed using resampling combined with the BPR loss equation, which incurs the same per-instance time complexity as the standard BPR loss computation. Although this effectively doubles the number of BPR loss computations, the constant factor is absorbed in the Big-O notation. Therefore, the overall time complexity of our method remains:
\begin{equation}
	2 \times O(2|E|Ld) = O(4|E|Ld) = O(|E|Ld).\label{eq:time_complexity}
\end{equation}
This demonstrates that our approach has the same asymptotic time complexity as the original LightGCN model. However, we acknowledge that in practice the additional BPR evaluation introduces an approximately twofold increase in the runtime of the loss computation, even though it does not change the asymptotic order.


\section{Experiments}
\label{sec:experiments}

This section presents the experimental evaluation of our proposed methods. First, we describe the experimental settings, including the baselines, datasets, and evaluation metrics employed in our study. Next, we report and analyze the experimental results, comparing our methods against strong baselines with respect to both accuracy and fairness-related metrics.

\subsection{Experimental setup}

To evaluate the effectiveness of our proposed methods, we conducted a series of experiments using well-established datasets, competitive baseline models, and widely adopted evaluation metrics. This subsection outlines the experimental setup, providing detailed descriptions of the baseline methods used for comparison, the datasets on which the experiments were performed, and the metrics employed to assess performance from both accuracy and fairness perspectives.

\subsubsection{Baselines}
To assess both the effectiveness and generality of our proposed method, we integrate it into several strong baseline models and compare their performance before and after applying our loss function. Specifically, we select LightGCN alongside three recently developed graph-based collaborative filtering methods that address fairness by either mitigating popularity bias or managing distribution shifts from different perspectives. These baselines embody diverse and competitive strategies for enhancing fairness in recommender systems. In the following, we briefly introduce each of these baselines.

\paragraph{LightGCN}
He et al. \cite{DBLP:conf/sigir/0001DWLZ020} introduced a simplified graph-based collaborative filtering model that relies solely on neighborhood aggregation, without any feature transformation or nonlinear activation. The core idea is to represent each user and item as the weighted sum of their neighbors' embeddings across multiple graph convolution layers. The final embeddings for users and items are then obtained by summing the embeddings from all layers, including the initial embeddings:
\begin{equation}
	\mathbf{e}_u = \sum_{k=0}^{K} \alpha_k \, \mathbf{e}_u^{(k)}, \quad
	\mathbf{e}_i = \sum_{k=0}^{K} \alpha_k \, \mathbf{e}_i^{(k)}. \label{eq:lgn_fembed}
\end{equation}
In Equation~\ref{eq:lgn_fembed}, \(\mathbf{e}_u^{(k)}\) and \(\mathbf{e}_i^{(k)}\) denote the user and item embeddings at the \(k\)-th layer, respectively, while \(\alpha_k\) is the weight assigned to that layer (typically uniform). Here, \(K\) represents the total number of propagation layers.

The layer-wise embedding propagation is defined as follows:
\begin{equation}
	\mathbf{e}_u^{(k)} = \sum_{i \in \mathcal{N}_u} \frac{1}{\sqrt{|\mathcal{N}_u|\,|\mathcal{N}_i|}} \, \mathbf{e}_i^{(k-1)}, \quad
	\mathbf{e}_i^{(k)} = \sum_{u \in \mathcal{N}_i} \frac{1}{\sqrt{|\mathcal{N}_u|\,|\mathcal{N}_i|}} \, \mathbf{e}_u^{(k-1)}. \label{eq:lgn_lembed}
\end{equation}
In Equation~\ref{eq:lgn_lembed}, \(\mathcal{N}_u\) and \(\mathcal{N}_i\) denote the sets of neighboring items for user \(u\) and neighboring users for item \(i\), respectively.

\paragraph{r-AdjNorm}
Zhao et al. \cite{DBLP:conf/sigir/0002WLCZDWSLW22} investigated the trade-off between accuracy and novelty in graph-based collaborative filtering models. Their study shows that symmetric neighborhood aggregation---commonly used in existing graph-based recommenders---can exacerbate popularity bias as the graph propagation depth increases. To counter this issue, they proposed r-AdjNorm, a simple yet effective plugin that adjusts the normalization strength in neighborhood aggregation layers to balance recommendation accuracy and novelty. This method can be seamlessly integrated into popular graph-based models like LightGCN without incurring additional computational cost. By introducing a learnable parameter \(r\), r-AdjNorm modifies the standard LightGCN aggregation as follows:
\begin{equation}
	\mathbf{e}_u^{(k)} = \sum_{i \in \mathcal{N}_u} \frac{1}{|\mathcal{N}_u|^r\,|\mathcal{N}_i|^{1-r}} \, \mathbf{e}_i^{(k-1)}, \quad
	\mathbf{e}_i^{(k)} = \sum_{u \in \mathcal{N}_i} \frac{1}{|\mathcal{N}_u|^r\,|\mathcal{N}_i|^{1-r}} \, \mathbf{e}_u^{(k-1)}. \label{eq:ran_lembed}
\end{equation}
Here, \(r \in [0,1]\) is a hyperparameter that controls the balance of normalization between users and items; when \(r = 0.5\), the symmetric normalization used in LightGCN is recovered.

\paragraph{Adaptive Popularity Debiasing Aggregator (APDA)}
Zhou et al. \cite{DBLP:conf/sigir/ZhouCDZZ023} addressed popularity bias in graph-based collaborative filtering models by mitigating the bias directly during the aggregation process. Unlike previous approaches that modify the loss function---thus risking the propagation of bias through the aggregation---APDA adaptively learns per-edge weights to counteract the influence of item popularity, based on a novel inverse popularity score. This debiasing is further enhanced by a weight scaling mechanism and residual connections, which help prevent over-smoothing. APDA can be seamlessly integrated with popular backbones like LightGCN, leading to notable improvements in both recommendation accuracy and fairness.

\paragraph{PopGo}
Zhang et al. \cite{DBLP:journals/tois/ZhangMZWC24} proposed a robust debiasing strategy for collaborative filtering models to combat popularity distribution shifts. They observed that user and item representations often learn spurious "popularity shortcuts" that impair generalization to out-of-distribution (OOD) data. To address this, PopGo introduces an interaction-wise shortcut modeling approach. It first trains a shortcut model to capture popularity-only patterns and quantifies a shortcut degree for each user-item pair. The main recommender model is then adjusted based on these degrees to emphasize popularity-agnostic information.

PopGo reduces reliance on popularity shortcuts through a two-stage process. First, the shortcut model is trained to capture the inherent popularity patterns using the following loss:
\begin{equation}
	\mathcal{L}_b = - \sum_{(u,i) \in \mathcal{O}^+} \log \left( \frac{e^{\beta_{ui}/\tau}}{\sum_{j \in \mathcal{N}_u^+} e^{\beta_{uj}/\tau}} \right), \label{eq:popgo_lb}
\end{equation}
where \(\beta_{ui}\) denotes the shortcut degree for interaction \((u,i)\), \(\tau\) is the temperature hyperparameter, \(\mathcal{O}^+\) is the set of observed (positive) interactions, and \(\mathcal{N}_u^+\) represents the set of positive neighbors for user \(u\).
Next, the debiased collaborative filtering model is optimized by incorporating the shortcut degrees into the prediction as follows:
\begin{equation}
	\mathcal{L} = - \sum_{(u,i) \in \mathcal{O}^+} \log \left( \frac{e^{\alpha_{ui} \cdot \beta_{ui}^* / \tau}}{\sum_{j \in \mathcal{N}_u^+} e^{\alpha_{uj} \cdot \beta_{uj}^* / \tau}} \right), \label{eq:popgo_lf}
\end{equation}
where \(\alpha_{ui}\) is the main model's score for user-item pair \((u,i)\) and \(\beta_{ui}^*\) is the shortcut degree learned by the shortcut model. The temperature hyperparameter \(\tau\) modulates the smoothness of the softmax distribution.

This two-stage process effectively mitigates the reliance on popularity shortcuts, thereby enhancing both the robustness and fairness of the recommender system.

\subsubsection{Datasets}
Three public real-world datasets are used to evaluate the performance of the proposed method: Epinions\footnote{\url{https://snap.stanford.edu/data/soc-Epinions1.html}}, iFashion\footnote{\url{https://github.com/wenyuer/POG}}, and MovieLens\footnote{\url{https://grouplens.org/datasets/movielens/}}. In line with our problem definition---and given that implicit feedback datasets are among the most widely used in recommender systems research \cite{DBLP:conf/sigir/0002WLCZDWSLW22}---these datasets capture implicit interactions between users and items.

\textbf{Epinions} \cite{DBLP:conf/sigir/ZhouCDZZ023} is a user-item interaction dataset collected from a consumer review platform. Here, user reviews are treated as implicit positive feedback, resulting in a sparse interaction matrix with noticeable popularity bias.
\textbf{iFashion} \cite{DBLP:conf/kdd/ChenHXGGSLPZZ19} is a real-world fashion recommendation dataset characterized by a strong long-tail distribution. It captures various user behaviors---such as clicks and purchases---thereby providing a challenging benchmark for studying fairness in recommendations.
\textbf{MovieLens} \cite{DBLP:conf/recsys/RashedGS19} is a widely used benchmark dataset containing user ratings for movies. Following common practice, ratings are converted into implicit feedback by selecting those with ratings of 4 or higher and treating them as positive interactions.

In Table~\ref{tbl:datasets}, we report the datasets' statistics, including the number of users, number of items, number of implicit feedback instances (or user-item interactions), and the density.

\begin{table}
	\renewcommand{\arraystretch}{1.2}
	\centering
	\caption{Statistics of the datasets.\label{tbl:datasets}}
	\begin{tabular}{ l | c c c c }
		\hline
		Dataset & \# users & \# items & \# interactions & Density \\
		\hline
		Epinions & 11,496 & 11,656 & 327,942 & 0.245\% \\
		iFashion & 23,405 & 24,803 & 378,713 & 0.065\% \\
		MovieLens & 887 & 824 & 52,781 & 7.221\% \\
		\hline
	\end{tabular}
\end{table}

\paragraph{Train-test split}
To evaluate the models, we adopt a common data splitting strategy widely used in recommender system research. For each user, 20\% of their interactions are randomly selected and held out as the test set, while the remaining 80\% are used for training. The held-out interactions remain completely hidden during the training phase, simulating realistic recommendation scenarios. This user-wise random holdout protocol aligns with the experimental settings used in LightGCN \cite{DBLP:conf/sigir/0001DWLZ020}, r-AdjNorm \cite{DBLP:conf/sigir/0002WLCZDWSLW22}, and APDA \cite{DBLP:conf/sigir/ZhouCDZZ023}, ensuring a fair and comparable evaluation across all methods.

\subsubsection{Evaluation metrics}
We evaluate model performance using two groups of metrics. The first group comprises fairness metrics, which assess the models' ability to mitigate popularity bias and provide balanced recommendations. The second group consists of standard accuracy and ranking metrics commonly employed in recommender system evaluations. Below, we detail the metrics used in each group.

\paragraph{Accuracy and ranking metrics}
In this study, we use the F1-score, NDCG, and MAP as accuracy and ranking metrics, all of which have values between 0 and 1, with the larger value being more desirable. Below, we provide a brief explanation of each metric.
\begin{itemize}
	
	\item \textbf{F1 Score:} The F1 Score is the harmonic mean of precision and recall, providing a balanced evaluation of a model's performance by simultaneously considering both the accuracy and the completeness of the recommendations. Precision is the proportion of recommended items that are relevant, while recall is the proportion of relevant items that have been successfully recommended. The F1 Score is defined as:
	\begin{equation}
		\text{F1Score} = 2 \times \frac{\text{Precision} \times \text{Recall}}{\text{Precision} + \text{Recall}}.
	\end{equation}

	\item \textbf{Normalized Discounted Cumulative Gain (NDCG):} NDCG is a ranking metric that evaluates the quality of the recommendations by comparing the proposed ranking with an ideal ranking, in which all relevant items are ranked at the top. It is computed by normalizing the Discounted Cumulative Gain (DCG) with the Ideal Discounted Cumulative Gain (IDCG), ensuring that the metric ranges between 0 and 1. For the top-\(k\) recommended items, NDCG is defined as:
	\begin{equation}
		\text{NDCG@}k = \frac{\text{DCG@}k}{\text{IDCG@}k}, \label{eq:ndcg}
	\end{equation}
	where
	\begin{align}
		\text{DCG@}k &= \sum_{i=1}^{k} \frac{2^{r(i)}-1}{\log_{2}(i+1)}, \label{eq:dcg} \\
		\text{IDCG@}k &= \sum_{i=1}^{REL_k} \frac{2^{r(i)}-1}{\log_{2}(i+1)}. \label{eq:idcg}
	\end{align}
	In these equations, \(r(i)\) denotes the relevance score of the item ranked at position \(i\), where typically a relevant item is assigned a score of 1 and an irrelevant item a score of 0. \(REL_k\) represents the ideal (i.e., perfectly sorted) list of the top-\(k\) most relevant items in the test set.

	\item \textbf{Mean Average Precision (MAP):} MAP measures the ranking quality of the recommender system by averaging the Average Precision (AP) scores across all users. For a given user \(u\), the AP for the top-\(k\) items is calculated as follows:
	\begin{equation}
		AP@k = \frac{1}{\text{number of relevant items}} \sum_{i=1}^{k} P@i \cdot r(i), \label{eq:ap}
	\end{equation}
	where \(P@i\) is the precision at the \(i\)-th position in the recommendation list, and \(r(i)\) is an indicator function that takes the value 1 if the item at rank \(i\) is relevant, and 0 otherwise. The overall MAP is then computed as the mean of the AP scores across all users.
	
\end{itemize}

\paragraph{Fairness metrics}
Since our primary focus in this research is on mitigating popularity bias, we adopt popularity bias metrics as our fairness criteria. Zhou et al. \cite{DBLP:conf/wsdm/Zhu0ZZWC21} introduced two metrics for quantifying popularity bias from both the item and user perspectives. Similar to conventional recommender system metrics such as NDCG, these popularity bias criteria are computed based on test cases. In this work, we utilize these two metrics to evaluate the impact of PBiLoss on recommender systems:
\begin{itemize}
	\item \textbf{Popularity-Rank Correlation for Users (PRU):} Evaluates the ability to balance popularity and ranking within each user's recommendation list by computing the correlation between the popularity scores and the corresponding ranks of the recommended items. It is defined as:
	\begin{equation}
		PRU = -\frac{1}{|U|} \sum_{u \in U} SRC\left( pop\left(\tilde{O}_{u}^{+}\right), \, rank_{u}\left(\tilde{O}_{u}^{+}\right) \right), \label{eq:pru}
	\end{equation}
	
	\item \textbf{Popularity-Rank Correlation for Items (PRI):} Assesses the balance between item popularity and its average ranking across all users. Specifically, it computes the correlation between an item's intrinsic popularity and the average rank it receives in the recommendation lists that include it. This is formulated as:
	\begin{equation}
		PRI = -\frac{1}{|I|} \sum_{i \in I} SRC\left( pop(i), \, \frac{1}{|U_i|}\sum_{u \in U_{i}} rank_{u}(i) \right), \label{eq:pri}
	\end{equation}
	In Equations \ref{eq:pru} and \ref{eq:pri}, the following notations are used:
	\begin{itemize}
		\item \(\tilde{O}_{u}^{+}\) represents the set of recommended items for user \(u\) drawn from the held-out (test) set.
		\item \(SRC(\cdot,\cdot)\) denotes the Spearman Rank Correlation between two sequences.
		\item \(rank_u(i)\) indicates the model's predicted rank of item \(i\) in the recommendation list for user \(u\).
		\item \(U_{i}\) is the set of users for whom item \(i\) appears in their test rankings.
		\item \(pop(i)\) specifies the popularity score of item \(i\), operationalized as its node degree.
	\end{itemize}
\end{itemize}

Both metrics assess how effectively a recommender system mitigates popularity bias. Being correlation-based, they evaluate the relationship between two variable sequences and yield values in the range \([-1, +1]\); values closer to 0 indicate a more balanced or fair treatment regarding popularity. Specifically, \(PRU\) measures the bias in the positions of items within individual user recommendation lists, while \(PRI\) evaluates the fairness of item exposure across all users. A high \(PRU\) value suggests that individual users' preferences have little influence on the ranking of recommended items, whereas a high \(PRI\) value indicates that unpopular items have limited opportunities to be recommended. In general, as these metric values approach 0, the recommender system is considered to be performing better in terms of promoting fairness.

\subsubsection{Parameter settings}
For all experiments, we adopt a LightGCN backbone comprising 4 graph convolutional layers and an embedding dimension of 64. The batch size is set to 1024 based on resource considerations and remains fixed across experiments to ensure consistency. An early stopping mechanism based on validation performance is employed to prevent overfitting. We used Late Learning Rate Decay \cite{DBLP:conf/aistats/Ren0Y24} in training our models. The models are initially trained with a learning rate of 0.001 for the first 50 epochs; thereafter, an exponential learning rate decay is applied, gradually reducing the rate to a minimum of 0.0001.

For hyperparameter tuning, the fairness regularization weight \(w\) and the balancing coefficient \(\alpha\) were optimized using grid search within a predefined range for each variant of our methodology. The selection criteria were based on the best validation performance and stable convergence behavior. This ensures that each variant operates under its most effective configuration while maintaining a consistent tuning procedure across all experiments.

\subsection{Experimental results}
This subsection presents the outcomes of our experiments and compares the performance of the proposed methods against several baseline models built on LightGCN. Our method, PBiLoss, is integrated as an add-on component and evaluated in four distinct variants:
\begin{enumerate}
	\item Popular positive with no popularity threshold (\emph{PopPos-NT})
	\item Popular positive with fixed popularity threshold (\emph{PopPos-FT})
	\item Popular negative with no popularity threshold (\emph{PopNeg-NT})
	\item Popular negative with fixed popularity threshold (\emph{PopNeg-FT})
\end{enumerate}

For each baseline model, we integrate all four variants of PBiLoss and analyze their impact on recommendation performance. Each subsection is dedicated to a specific baseline model, presenting results both with and without the integration of the PBiLoss variants. Evaluation outcomes are reported across multiple metrics, demonstrating that PBiLoss not only enhances fairness in terms of mitigating popularity bias but also maintains competitive overall performance.

\subsubsection{Integration with LightGCN}

We first evaluate our approach using LightGCN \cite{DBLP:conf/sigir/0001DWLZ020}, which serves as one of our primary baselines. Each of the four PBiLoss variants is integrated with LightGCN, and the resulting performance is presented below. This comparison highlights the impact of our approach on both fairness and overall recommendation performance.

\begin{table}
	\renewcommand{\arraystretch}{1.2}
	\centering
	\caption{Experimental results of the PBiLoss variants integrated with LightGCN. Symbols $\uparrow$ and $\downarrow$ indicate that higher and lower values are better, respectively. The best and second-best results are highlighted in \textbf{bold} and \underline{underlined}, respectively.\label{tbl:lgn_res}}
	\scalebox{0.8}{
		\begin{tabular}{ c | l | c c c c c }
			\hline
			Dataset & Metrics & LightGCN \cite{DBLP:conf/sigir/0001DWLZ020} & +PopPos-NT & +PopPos-FT & +PopNeg-NT & +PopNeg-FT \\
			\hline
			\multirow{5}{*}{Epinions}
			& PRU $\downarrow$ & 0.5415 & \underline{0.5191} & 0.5357 & \textbf{0.5187} & 0.5222 \\
			& PRI $\downarrow$ & 0.5327 & \textbf{0.5066} & 0.5266 & \underline{0.5076} & 0.5123 \\
			\cdashline{2-7}
			& F1@10 $\uparrow$ & 0.0458 & 0.0459 & \textbf{0.0467} & 0.0461 &\underline{0.0465} \\
			& NDCG@10 $\uparrow$ & 0.0666 & 0.0667 & \textbf{0.0673} & 0.0669 & \underline{0.0671} \\
			& MAP@10 $\uparrow$ & 0.0295 & 0.0297 & \textbf{0.0299} & \underline{0.0298} & 0.0297 \\
			\hline
			\multirow{5}{*}{iFashion}
			& PRU $\downarrow$ & 0.5094 & 0.5075 & 0.5062 & \underline{0.5057} & \textbf{0.4901} \\
			& PRI $\downarrow$ & 0.5748 & 0.5741 & 0.5705 & \underline{0.5702} & \textbf{0.5632} \\
			\cdashline{2-7}
			& F1@10 $\uparrow$ & 0.0252 & 0.0251 & \underline{0.0257} & 0.0252 & \textbf{0.0258} \\
			& NDCG@10 $\uparrow$ & 0.0391 & 0.0389 & \underline{0.0399} & 0.0391 & \textbf{0.0401} \\
			& MAP@10 $\uparrow$ & 0.0202 & 0.0201 & \underline{0.0205} & 0.0201 & \textbf{0.0207} \\
			\hline
			\multirow{5}{*}{MovieLens}
			& PRU $\downarrow$ & 0.5678 & 0.5468 & \textbf{0.5405} & \underline{0.5453} & 0.5484 \\
			& PRI $\downarrow$ & 0.8159 & 0.8047 & \textbf{0.8025} & \underline{0.8032} & 0.8043 \\
			\cdashline{2-7}
			& F1@10 $\uparrow$ & 0.2004 & 0.2041 & \underline{0.2058} & 0.2038 & \textbf{0.2062} \\
			& NDCG@10 $\uparrow$ & 0.3075 & 0.3146 & \underline{0.3158} & 0.3144 & \textbf{0.3164} \\
			& MAP@10 $\uparrow$ & 0.1286 & 0.1329 & \underline{0.1341} & 0.1331 & \textbf{0.1343} \\
			\hline
		\end{tabular}
	}
\end{table}

In Table \ref{tbl:lgn_res}, the experimental results demonstrate that integrating PBiLoss with LightGCN consistently enhances performance across all fairness, accuracy, and ranking metrics. Notably, both the Popular Positive (PopPos) and Popular Negative (PopNeg) strategies lead to reductions in the popularity bias metrics (PRU and PRI) while simultaneously improving recommendation performance metrics such as F1-Score, NDCG, and MAP. Among the fairness metrics, PRU---representing user-side popularity bias---shows the most substantial reduction, highlighting the effectiveness of the proposed loss function in balancing user-level exposure to popular items. This observation suggests that PBiLoss has a particularly strong impact on enhancing fairness from the user perspective.

In addition, the PopNeg-FT variant frequently achieves the best overall balance, demonstrating the strongest performance in reducing bias and improving ranking quality in several cases. These improvements are especially pronounced on the MovieLens dataset, where the proposed methods clearly outperform the base model.

The strong performance of PopNeg-FT can be attributed to how fixed threshold popularity modeling interacts with Popular Negative (PopNeg) strategy in pairwise ranking optimization. In BPR-based GNN recommenders, model updates are largely driven by negative samples, and popularity bias is reinforced when highly connected (popular) items are insufficiently penalized. PopNeg-FT explicitly separates popular and unpopular items using a fixed threshold, ensuring that highly popular items are consistently treated as hard negatives when they are irrelevant to a user. This deterministic separation provides a stable and interpretable signal that prevents popularity information from being diluted by the long-tailed degree distribution, which can occur in probabilistic sampling (No popularity threshold). In contrast, probabilistic sampling may still under-penalize popular items due to smoothing effects in the degree-based distribution. As a result, PopNeg-FT more effectively suppresses the over-ranking of popular items while preserving ranking accuracy, leading to a stronger and more consistent fairness and accuracy trade-off across datasets.

Although the gains in accuracy and ranking metrics are modest compared to the improvements in fairness metrics, the fact that PBiLoss enhances fairness without sacrificing accuracy underscores its practical value. The robust performance of both the PopPos and PopNeg strategies---across different sampling methods---further reinforces the flexibility and robustness of the proposed approach. Overall, these results validate the effectiveness of PBiLoss in promoting fairer recommendations while maintaining competitive accuracy.


\subsubsection{Integration with r-AdjNorm}

\begin{table}
	\renewcommand{\arraystretch}{1.2}
	\centering
	\caption{Experimental results of PBiLoss variants integrated with r-AdjNorm.\label{tbl:adjnorm_res}}
	\scalebox{0.8}{
		\begin{tabular}{ c | l | c c c c c }
			\hline
			Dataset & Metrics & r-AdjNorm \cite{DBLP:conf/sigir/0002WLCZDWSLW22} & +PopPos-NT & +PopPos-FT & +PopNeg-NT & +PopNeg-FT \\
			\hline
			\multirow{5}{*}{Epinions}
			& PRU $\downarrow$ & 0.5272 & 0.5351 & 0.5352 & \underline{0.5325} & \textbf{0.5104} \\
			& PRI $\downarrow$ & 0.5539 & 0.5547 & 0.5547 & \underline{0.5533} & \textbf{0.5414} \\
			\cdashline{2-7}
			& F1@10 $\uparrow$ & 0.0462 & 0.0460 & 0.0461 & \underline{0.0462} & \textbf{0.0463} \\
			& NDCG@10 $\uparrow$ & 0.0664 & 0.0661 & 0.0663 & \underline{0.0665} & \textbf{0.0669} \\
			& MAP@10 $\uparrow$ & 0.0292 & 0.0291 & \underline{0.0293} & 0.0292 & \textbf{0.0297} \\
			\hline
			\multirow{5}{*}{iFashion}
			& PRU $\downarrow$ & 0.5248 & 0.5232 & \underline{0.5136} & 0.5223 & \textbf{0.4949} \\
			& PRI $\downarrow$ & 0.6305 & 0.6301 & \underline{0.6292} & 0.6295 & \textbf{0.6201} \\
			\cdashline{2-7}
			& F1@10 $\uparrow$ & 0.0248 & 0.0247 & 0.0247 & \underline{0.0248} & \textbf{0.0249} \\
			& NDCG@10 $\uparrow$ & 0.0383 & 0.0382 & 0.0382 & \underline{0.0383} & \textbf{0.0385} \\
			& MAP@10 $\uparrow$ & 0.0196 & 0.0196 & 0.0195 & \underline{0.0197} & \textbf{0.0198} \\
			\hline
			\multirow{5}{*}{MovieLens}
			& PRU $\downarrow$ & 0.5572 & 0.5601 & 0.5551 & \textbf{0.5004} & \underline{0.5355} \\
			& PRI $\downarrow$ & 0.8068 & 0.8120 & 0.8062 & \textbf{0.7699} & \underline{0.7932} \\
			\cdashline{2-7}
			& F1@10 $\uparrow$ & 0.2038 & 0.2008 & \underline{0.2051} & 0.2039 & \textbf{0.2057} \\
			& NDCG@10 $\uparrow$ & 0.3124 & 0.3081 & \textbf{0.3136} & 0.3074 & \underline{0.3132} \\
			& MAP@10 $\uparrow$ & 0.1313 & 0.1291 & \underline{0.1315} & 0.1288 & \textbf{0.1318} \\
			\hline
		\end{tabular}
	}
\end{table}

Table \ref{tbl:adjnorm_res} presents the experimental results of integrating the PBiLoss variants with the r-AdjNorm method \cite{DBLP:conf/sigir/0002WLCZDWSLW22}. Overall, the PBiLoss variants continue to demonstrate their effectiveness. The improvements in user-side fairness (measured by PRU) are more pronounced than those on the item side (PRI), confirming that the proposed loss function is particularly effective at personalizing the recommendations for users. This trend is consistent across all datasets, with the PopNeg-FT variant frequently achieving the best performance, thereby underscoring its strength.

While the accuracy and ranking metrics exhibit only marginal changes, the consistent improvements across these metrics underscore the stability of PBiLoss even when integrated with more complex or recent models. This observation suggests that PBiLoss can serve as a general-purpose regularization technique for bias mitigation in various GNN-based recommender systems, offering enhanced fairness at minimal cost to overall performance.

\subsubsection{Integration with APDA}

\begin{table}
	\renewcommand{\arraystretch}{1.2}
	\centering
	\caption{Experimental results of PBiLoss variants integrated with APDA.\label{tbl:adpa_res}}
	\scalebox{0.8}{
		\begin{tabular}{ c | l | c c c c c }
			\hline
			Dataset & Metrics & APDA \cite{DBLP:conf/sigir/ZhouCDZZ023} & +PopPos-NT & +PopPos-FT & +PopNeg-NT & +PopNeg-FT \\
			\hline
			\multirow{5}{*}{Epinions}
			& PRU $\downarrow$ & 0.4969 & 0.4932 & 0.5047 & \underline{0.4909} & \textbf{0.4782} \\
			& PRI $\downarrow$ & 0.4282 & 0.4281 & 0.4371 & \underline{0.4256} & \textbf{0.4121} \\
			\cdashline{2-7}
			& F1@10 $\uparrow$ & 0.0493 & \underline{0.0497} & \textbf{0.0498} & 0.0496 & 0.0496 \\
			& NDCG@10 $\uparrow$ & 0.0716 & \textbf{0.0729} & 0.0723 & \underline{0.0728} & 0.0723 \\
			& MAP@10 $\uparrow$ & 0.0319 & \underline{0.0326} & 0.0323 & \textbf{0.0327} & 0.0323 \\
			\hline
			\multirow{5}{*}{iFashion}
			& PRU $\downarrow$ & 0.5375 & 0.5239 & \textbf{0.4992} & 0.5238 & \underline{0.5006} \\
			& PRI $\downarrow$ & 0.5430 & 0.5277 & \textbf{0.5066} & 0.5266 & \underline{0.5108} \\
			\cdashline{2-7}
			& F1@10 $\uparrow$ & 0.0288 & \underline{0.0291} & 0.0289 & \textbf{0.0297} & 0.0288 \\
			& NDCG@10 $\uparrow$ & 0.0439 & \underline{0.0448} & 0.0446 & \textbf{0.0456} & 0.0446 \\
			& MAP@10 $\uparrow$ & 0.0225 & 0.0228 & 0.0229 & \underline{0.0231} & \textbf{0.0232} \\
			\hline
			\multirow{5}{*}{MovieLens}
			& PRU $\downarrow$ & 0.5648 & 0.5652 & \textbf{0.5454} & 0.5742 & \underline{0.5569} \\
			& PRI $\downarrow$ & 0.7987 & 0.8006 & \underline{0.7918} & 0.7934 & \textbf{0.7876} \\
			\cdashline{2-7}
			& F1@10 $\uparrow$ & 0.2145 & \underline{0.2163} & \textbf{0.2171} & 0.2134 & 0.2154 \\
			& NDCG@10 $\uparrow$ & 0.3325 & 0.3334 & \underline{0.3349} & 0.3289 & \textbf{0.3344} \\
			& MAP@10 $\uparrow$ & 0.1415 & 0.1413 & \textbf{0.1426} & 0.1389 & \underline{0.1418} \\
			\hline
		\end{tabular}
	}
\end{table}

We then evaluate our method on Adaptive Popularity Debiasing Aggregator (APDA) \cite{DBLP:conf/sigir/ZhouCDZZ023}, a baseline designed to reduce popularity bias. By incorporating each PBiLoss variant, we assess whether further fairness gains can be achieved beyond APDA's built-in debiasing mechanisms.

Table \ref{tbl:adpa_res} demonstrates that integrating PBiLoss variants with APDA leads to improvements in fairness by reducing popularity bias, as evidenced by decreases in both PRU and PRI, with the PopNeg-FT variant exhibiting the strongest performance. However, the overall gains are subtler compared to other baseline models, likely because APDA already employs robust debiasing techniques to address popularity-driven signals. Despite this, accuracy and ranking metrics remain stable or exhibit slight improvements, indicating that PBiLoss can provide complementary fairness benefits without sacrificing recommendation quality.

\subsubsection{Integration with PopGo}

\begin{table}
	\renewcommand{\arraystretch}{1.2}
	\centering
	\caption{Experimental results of PBiLoss variants integrated with PopGo.\label{tbl:popgo_res}}
	\scalebox{0.8}{
		\begin{tabular}{ c | l | c c c c c }
			\hline
			Dataset & Metrics & PopGo \cite{DBLP:journals/tois/ZhangMZWC24} & +PopPos-NT & +PopPos-FT & +PopNeg-NT & +PopNeg-FT \\
			\hline
			\multirow{5}{*}{Epinions}
			& PRU $\downarrow$ & 0.6101 & \underline{0.6011} & 0.6136 & \textbf{0.5898} & 0.6144 \\
			& PRI $\downarrow$ & 0.5644 & \underline{0.5588} & 0.5679 & \textbf{0.5543} & 0.5682 \\
			\cdashline{2-7}
			& F1@10 $\uparrow$ & 0.0501 & \underline{0.0504} & 0.0497 & \textbf{0.0515} & 0.0493 \\
			& NDCG@10 $\uparrow$ & 0.0725 & \textbf{0.0738} & 0.0725 & \underline{0.0733} & 0.0721 \\
			& MAP@10 $\uparrow$ & 0.0325 & \textbf{0.0334} & 0.0321 & \underline{0.0329} & 0.0323 \\
			\hline
			\multirow{5}{*}{iFashion}
			& PRU $\downarrow$ & 0.6493 & 0.6490 & \underline{0.6316} & 0.6485 & \textbf{0.6166} \\
			& PRI $\downarrow$ & 0.6816 & 0.6815 & \underline{0.6700} & 0.6812 & \textbf{0.6577} \\
			\cdashline{2-7}
			& F1@10 $\uparrow$ & 0.0283 & 0.0284 & \textbf{0.0293} & 0.0283 & \underline{0.0288} \\
			& NDCG@10 $\uparrow$ & 0.0439 & 0.0439 & \textbf{0.0445} & 0.0440 & \underline{0.0442} \\
			& MAP@10 $\uparrow$ & 0.0227 & 0.0227 & \textbf{0.0234} & 0.0229 & \underline{0.0232} \\
			\hline
			\multirow{5}{*}{MovieLens}
			& PRU $\downarrow$ & 0.6701 & \textbf{0.6512} & 0.6649 & 0.6633 & \underline{0.6615} \\
			& PRI $\downarrow$ & 0.8458 & \underline{0.8352} & \textbf{0.8342} & 0.8457 & 0.8378 \\
			\cdashline{2-7}
			& F1@10 $\uparrow$ & 0.2037 & 0.2038 & 0.2039 & \textbf{0.2052} & \underline{0.2044} \\
			& NDCG@10 $\uparrow$ & 0.3107 & 0.3106 & \underline{0.3115} & 0.3107 & \textbf{0.3121} \\
			& MAP@10 $\uparrow$ & 0.1278 & \underline{0.1282} & 0.1276 & 0.1278 & \textbf{0.1293} \\
			\hline
		\end{tabular}
	}
\end{table}

Finally, we test the variants of our proposed method on PopGo \cite{DBLP:journals/tois/ZhangMZWC24}, a baseline that incorporates fairness in its learning process. The experimental results in Table \ref{tbl:popgo_res} show that integrating PBiLoss variants with PopGo leads to notable improvements in mitigating popularity bias. PopGo, originally designed to address a different aspect of fairness and built upon a separate evaluation framework, does not explicitly target popularity bias. Consequently, when used in isolation, it may not adequately mitigate this bias and, in many cases, may even exacerbate it. By complementing PopGo with PBiLoss, we achieve further fairness gains by effectively reducing popularity bias.

\subsection{Statistical significance analysis}

To alleviate concerns about statistical significance of performance gains, especially in NDCG, we performed a Friedman non-parametric test between the LightGCN baseline and our four variants on the three datasets. The Friedman statistic was \({\chi}^{2}\)(4) = 9.8305, p = 0.0434, rejecting the null hypothesis of no difference among the five models at \(\alpha\) = 0.05. This suggests significant statistical differences in NDCG rankings across datasets to support the observed improvements by our variants even if the actual gains may be small in absolute terms \cite{DBLP:journals/jmlr/Demsar06}.

\subsection{Discussion}

In most experiments, the PopNeg-FT variant outperformed the other PBiLoss variants in reducing popularity bias while maintaining stable or even enhanced accuracy and ranking performance. This consistent superiority indicates that applying negative sampling based on item popularity---with a controlled threshold---serves as a robust mechanism for penalizing the overexposure of popular items, thereby promoting more fair and balanced recommendations.

When comparing the variants, PopNeg methods generally exhibited stronger performance than their PopPos counterparts, particularly in fairness metrics. This can be attributed to the fact that popularity bias in recommender systems is often driven more by the overrepresentation of popular items as positive samples in the BPR loss. By explicitly targeting this overrepresentation through negative sampling, the PopNeg variants offer a more direct and effective debiasing mechanism. Nonetheless, PopPos methods still yield measurable improvements, contributing to the overall enhancement of fairness in the recommendations.

The dataset characteristics reveal important trends. For instance, the iFashion dataset, which is relatively sparse, exhibits better improvements---especially with the PopNeg variants---due to its lower density, which allows for more effective debiasing. In contrast, MovieLens, despite being dense, suffers from a strong concentration of popular items and an imbalanced exposure, rendering fairness improvements more challenging. Furthermore, across all datasets, employing a fixed popularity threshold method usually outperforms no popularity threshold method. The absence of a fixed threshold tends to be less effective, particularly in sparse environments where uncontrolled sampling may exacerbate existing biases. Also, when using the no popularity threshold method, the model might pronounce items that are excessively popular or unpopular, which can lead to unstable training dynamics and diminished fairness gains. In contrast, the fixed popularity threshold method enables finer control over which items are considered popular during training, resulting in more reliable performance improvements.

Overall, these findings highlight several critical insights. First, PBiLoss---especially the PopNeg-FT variant---emerges as a robust and effective method across diverse model architectures and datasets. Second, dataset density plays a pivotal role in determining the magnitude of fairness gains and in selecting the most appropriate variant. Finally, employing a fixed popularity threshold consistently enhances both fairness and performance stability, underscoring its value as a regularization strategy in debiasing.


\subsection{Hyperparameters study}

To evaluate the sensitivity and robustness of our proposed PBiLoss, we conduct a comprehensive hyperparameter study by varying key parameters that influence the training process. Specifically, we examine the regularization weight \(w\), which determines the strength of the PBiLoss component, and the fixed popularity threshold \(\alpha\), which is used to distinguish popular items in the fixed popularity threshold method.

\begin{figure}
	\centering
	\begin{subfigure}{0.45\textwidth}
		\includegraphics[width=\linewidth]{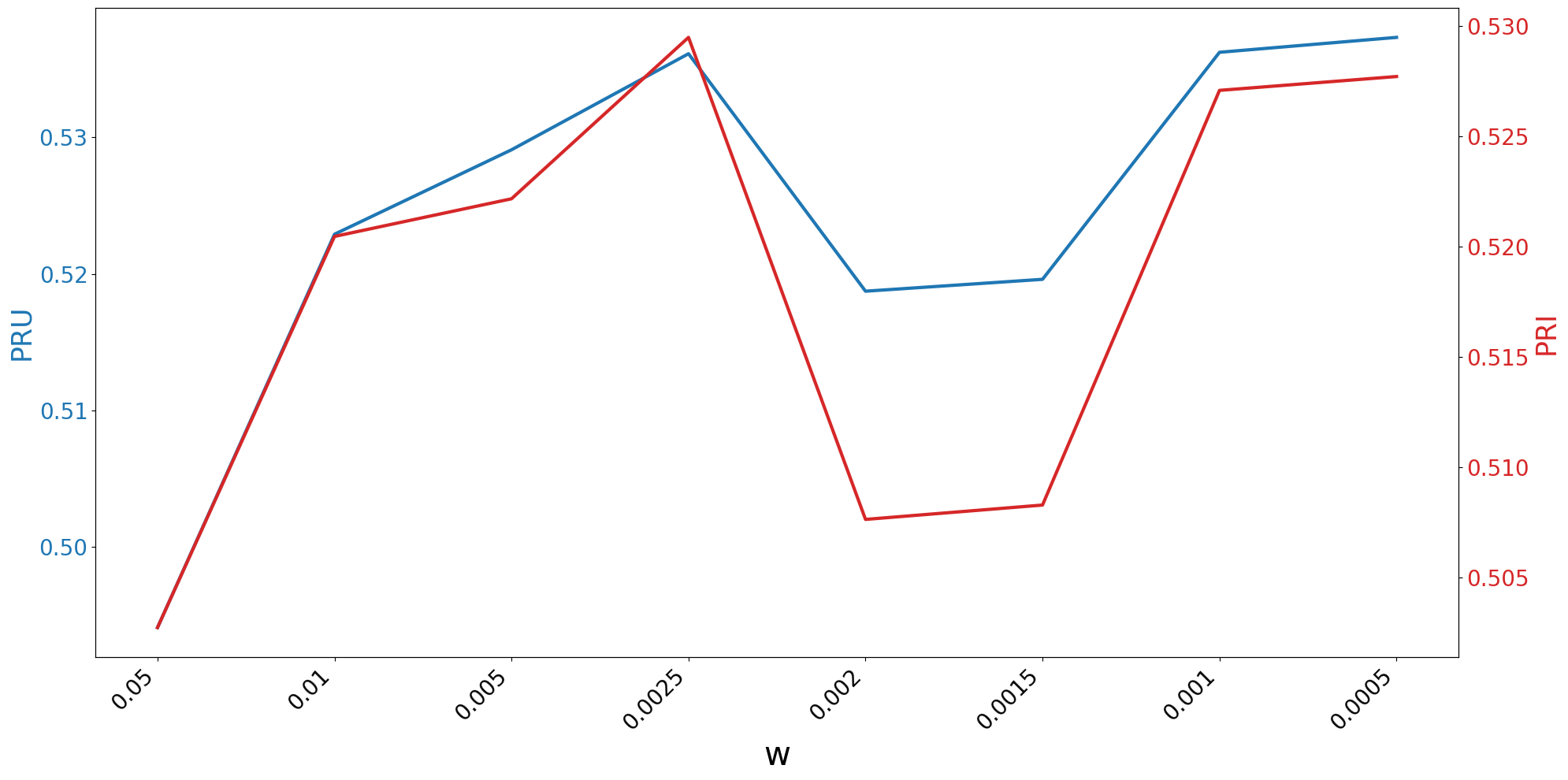}
	\end{subfigure}
	\begin{subfigure}{0.45\textwidth}
		\includegraphics[width=\linewidth]{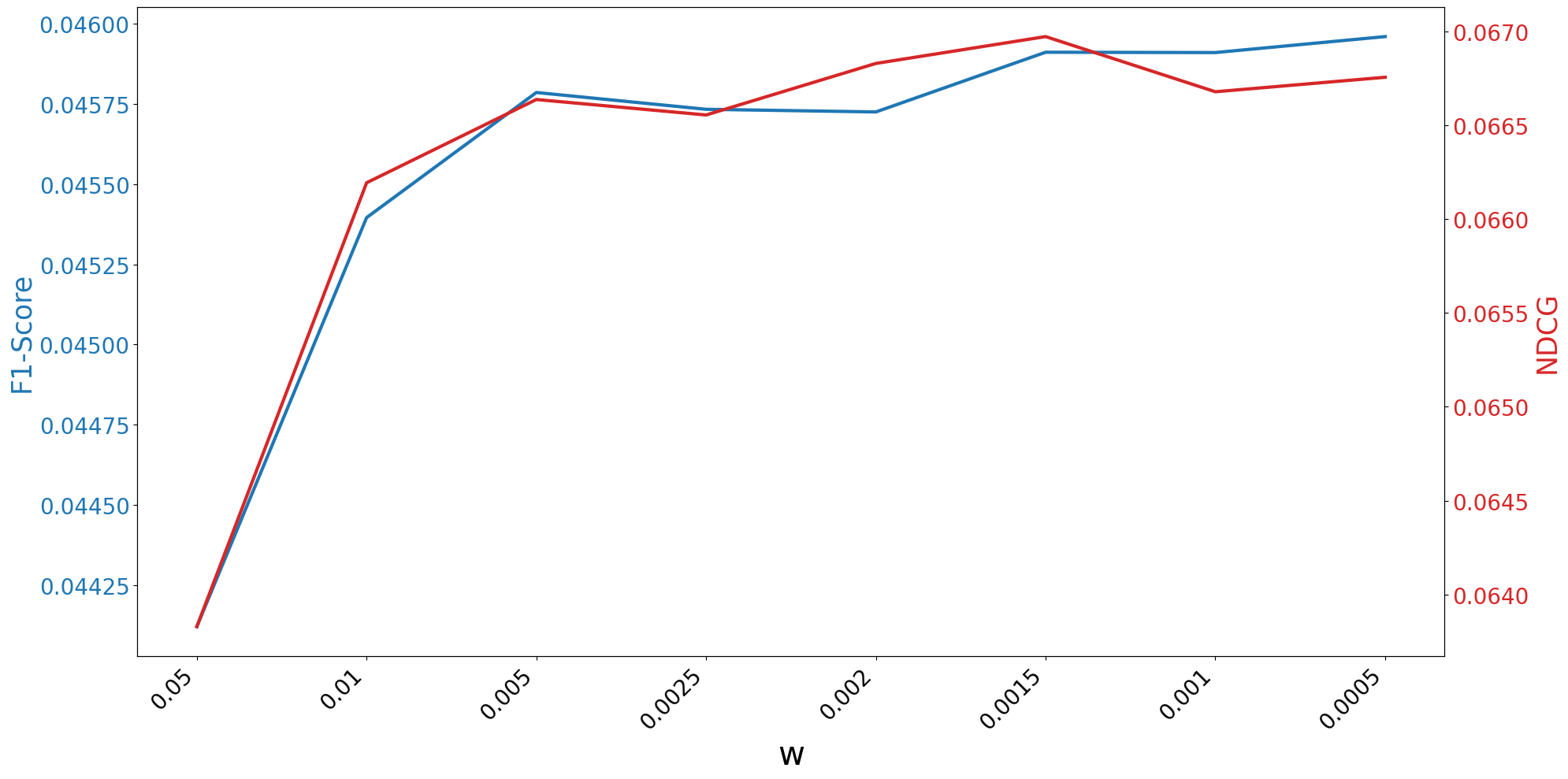}
	\end{subfigure}
	\caption{The effect of \(w\) on PBiLoss.}
	\label{fig:epinions_popnegnt}
\end{figure}

For the regularization weight \(w\), we explore a range of values from 0.05 to 0.0005 and examine their impact on both recommendation performance (e.g., NDCG, F1-Score) and fairness metrics (PRI and PRU). The results indicate that moderate values of \(w\) achieve an optimal balance between accuracy and fairness. Lower values of \(w\) result in minimal regularization, which fails to sufficiently reduce popularity bias and limits fairness improvements. Conversely, excessively high values lead to over-penalization, degrading recommendation accuracy.

\begin{figure}
	\centering
	\begin{subfigure}{0.45\textwidth}
		\includegraphics[width=\linewidth]{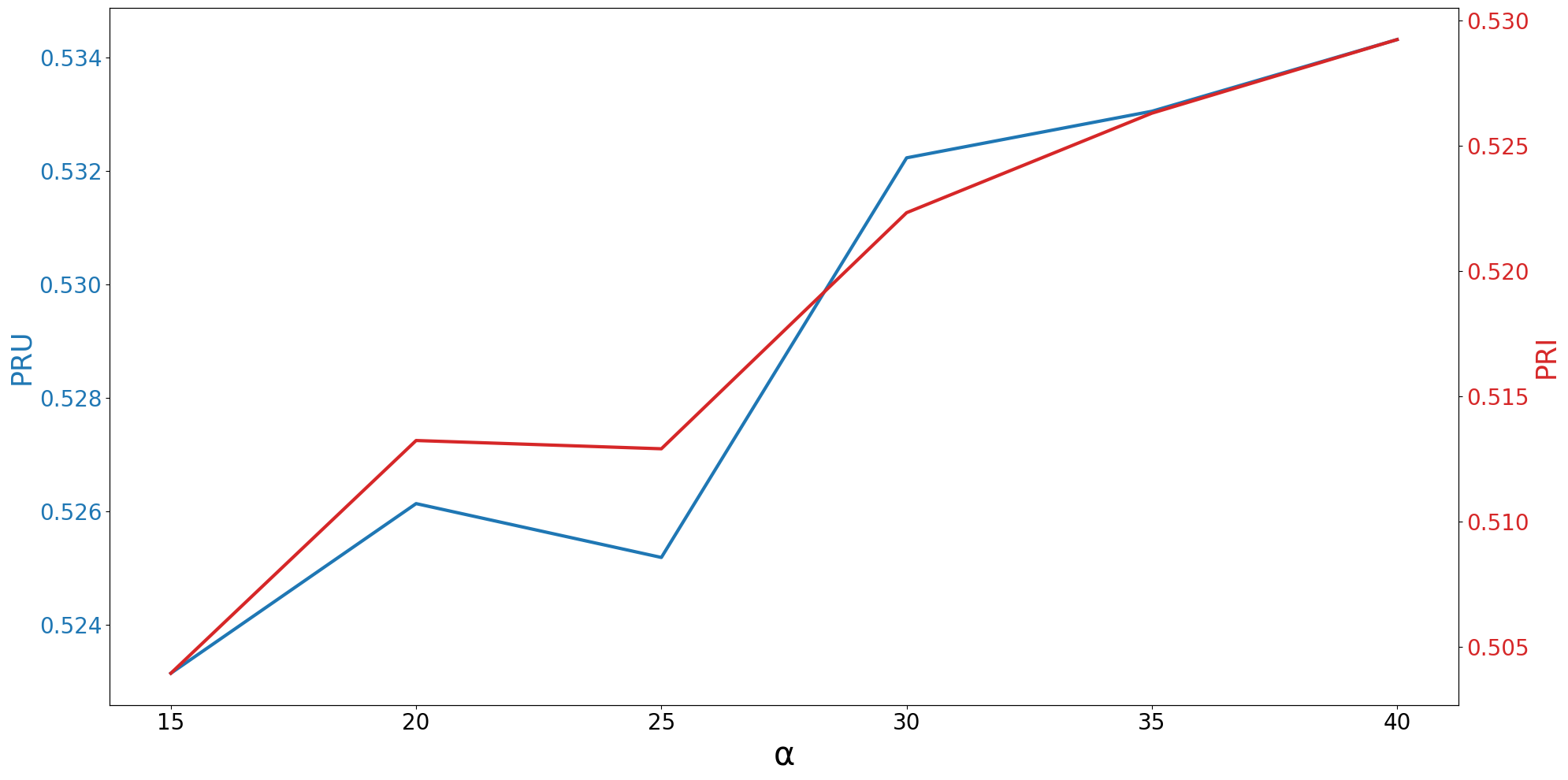}
	\end{subfigure}
	\begin{subfigure}{0.45\textwidth}
		\includegraphics[width=\linewidth]{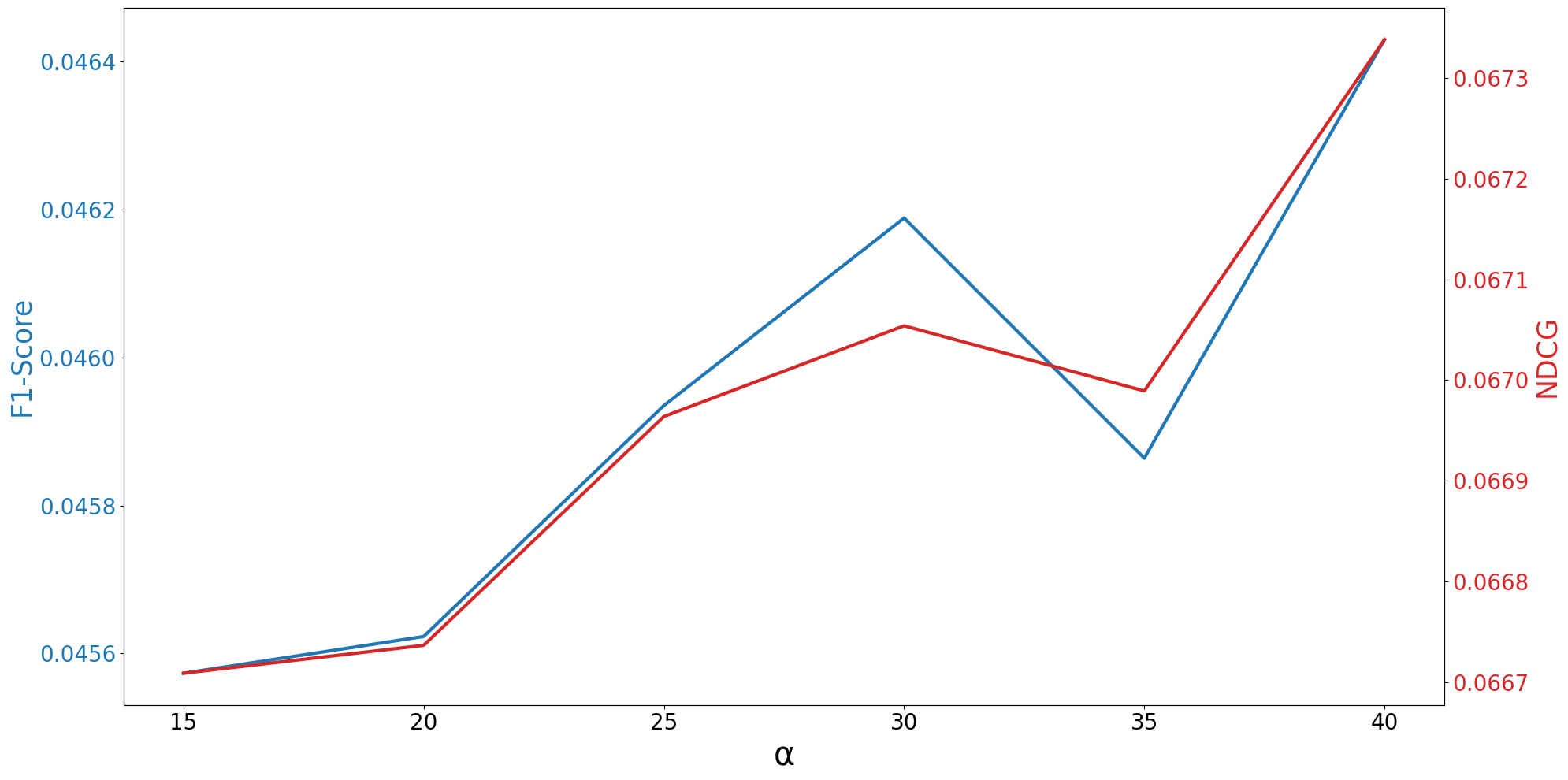}
	\end{subfigure}
	\caption{The effect of \(\alpha\) on PBiLoss with fixed popularity threshold.}
	\label{fig:epinions_popnegft_alpha}
\end{figure}

Similarly, we evaluate the impact of the popularity threshold \(\alpha\) used in our fixed popularity threshold method. We test fixed thresholds based on the percentiles of the most popular items (e.g., top 15\%, 20\%, and 25\%). Our experiments reveal that setting \(\alpha\) to designate approximately the top 20\% of items as popular leads to the most balanced improvements in both fairness and accuracy metrics. Overall, these observations confirm that PBiLoss is relatively robust over a sensible range of hyperparameter settings; however, careful tuning of \(w\) and \(\alpha\) can further enhance its ability to reduce popularity bias without significantly compromising recommendation effectiveness.



\section{Conclusion}
\label{sec:conclusion}

In this paper, we introduced PBiLoss, a novel loss regularization technique designed to mitigate popularity bias in graph-based recommender systems. Popularity bias remains a persistent challenge in recommendation tasks, often leading to the over-promotion of popular items at the expense of long-tail content and, consequently, reducing exposure diversity. To address this issue, we proposed two sampling strategies---PopPos and PopNeg---that integrate PBiLoss with the widely used BPR loss and rebalance user-item interactions during training by explicitly emphasizing unpopular items or de-emphasizing popular items.

We evaluated our approach on three benchmark datasets---Epinions, iFashion, and MovieLens---using a range of strong baseline models, including LightGCN, r-AdjNorm, APDA, and PopGo. The experimental results demonstrated that both the PopPos and PopNeg variants consistently reduce popularity bias while maintaining or even improving overall recommendation accuracy. Key metrics such as the Popularity-Rank correlation for Users (PRU) and Popularity-Rank correlation for Items (PRI) exhibited significant improvements, highlighting enhanced fairness and diversity in the recommendations. Overall, PBiLoss offers a model-agnostic, easily integrable, and practically effective solution for improving fairness in recommender systems under popularity bias.

\paragraph{Limitations}
Despite its effectiveness, our work has several limitations that open avenues for further research. First, PBiLoss relies on a specific popularity signal (currently degree-based), which may not fully capture context-dependent or temporal dynamics of item popularity. Second, our experiments focus on implicit-feedback, graph-based recommendation; extending the method to other paradigms (e.g., session-based, sequential, or multi-modal recommendation) may require non-trivial adaptations. Third, while we observe favorable trade-offs between fairness and accuracy, our analysis is restricted to a particular set of fairness and bias metrics, and a more comprehensive evaluation---including user-centric and group-fairness measures---remains an important direction.

\paragraph{Future works}
In future work, we plan to explore alternative popularity criteria beyond node degree---for instance, leveraging HITS (hub and authority) scores, temporal popularity trends, or content-aware popularity signals---to capture more nuanced aspects of item importance. Additionally, we aim to apply PBiLoss to broader recommendation settings and datasets, including industrial-scale platforms, to more rigorously assess its generalizability and robustness. Beyond loss-based regularization, a key direction is to design or modify model architectures so that they inherently mitigate popularity bias, potentially leading to more robust, interpretable, and equitable representations that harmonize accuracy, diversity, and fairness.


\section*{Data availability statement}
The datasets used in this study (Epinions \cite{DBLP:conf/sigir/ZhouCDZZ023}, iFashion \cite{DBLP:conf/kdd/ChenHXGGSLPZZ19}, and MovieLens \cite{DBLP:conf/recsys/RashedGS19}) are publicly available from their respective providers. 
The code used to generate the experimental results is publicly available at \url{https://github.com/MhmdNmi/PBiLoss}.

\bibliographystyle{plain} 
\bibliography{references}

\end{document}